\documentclass{article}

\usepackage{amsfonts,amssymb}
\usepackage{amsmath}
\usepackage{mathtools}
\usepackage{epsfig}
\usepackage{float}
\usepackage{indentfirst}
\usepackage{color}
\usepackage{caption}
\usepackage{graphicx, subfigure}
\usepackage{multirow}
\usepackage[table]{xcolor}
\usepackage{rotating}
\usepackage{ulem}
\usepackage{xcolor,cancel}
\usepackage{tikz}
\usetikzlibrary{shapes.misc}
\usetikzlibrary{shapes,snakes}
\usepackage{appendix}
\usepackage{comment}
\usepackage{yfonts}
\usepackage{mathrsfs}
\usepackage{rotating}

\makeatletter
  \@addtoreset{section}{part}
  \@addtoreset{@ppsaveapp}{part}
\makeatother

\numberwithin{equation}{section}

\begin{document}

\begin{center}
{\bf Active cloaking for clusters of pins in thin plates}
\end{center}

\begin{center}
J. O'Neill$^1$,  \"O. Selsil$^1$, S.G. Haslinger$^{1}$, N.V. Movchan$^1$ and R.V. Craster$^2$
\end{center}

\begin{center} 
$^1$ Department of Mathematical Sciences, University of Liverpool, Liverpool L69 7ZL, UK
\end{center}

\vspace{-0.8cm}

\begin{center}
$^2$ Department of Mathematical Sciences, Imperial College London, South Kensington Campus, London, SW7 2AZ, UK.
\end{center}

\begin{abstract}
This paper considers active cloaking of a square array of evenly spaced pins in a Kirchhoff plate in the presence of flexural waves. Active sources are distributed exterior to the cluster and are represented by the non-singular Green's function for the biharmonic operator. The complex amplitudes of the active sources, which cancel out selected multipole orders of the scattered field, are found by solving an algebraic system of equations. For frequencies in the zero-frequency stop band, we find that a small number of active sources located on a grid is sufficient for cloaking. For higher frequencies, we achieve efficient cloaking with the active sources positioned on a circle surrounding the cluster. We demonstrate the cloaking efficiency with several numerical illustrations, considering key frequencies from band diagrams and dispersion surfaces for a Kirchhoff plate pinned in a doubly periodic fashion.

\end{abstract}

\section{Introduction}
\label{Intro}

Following the seminal papers by Leonhardt \cite{UL} and Pendry {\it et al.} \cite{JBP_DS_DRS} in 2006, there has been great interest in the topic of cloaking, i.e. the elimination and control of scattering of waves by objects. There is extensive research on cloaking by diverting incident waves through a region exterior to an object. The object is shielded from the incident field rendering it invisible to an observer exterior to the cloaking region.
This method of cloaking has been developed for acoustics, electromagnetism, elasticity and water waves, see for example \cite{GWM_MB_JRW}--\cite{DJC_ISJ_NVM_ABM_MB_RCM}. These are examples of passive cloaks, where properties of the cloak are designed such that cloaking is achieved for a particular set of conditions; for changes in frequency and type of incident field, the cloak has to be redesigned. There also exists complementary research into active cloaking - a technique that involves several active devices located exterior to an object, which produce fields that cancel scattering from the object and  produce small scattered fields themselves. The active devices are able to adapt to changes in both frequency and type of incident field. Active cloaking techniques do not have the  drawbacks, discussed by Monticone \& Al\`u \cite{Alu_DoCloaks} and Chen {\it et al.} \cite{P-YC_CA_AA}, associated with their passive counterparts. The method of active cloaking was investigated in the context of the two-dimensional Laplace and Helmholtz equations by Guevara Vasquez {\it et al.} \cite{GVF_GWM_DO_2009a}--\cite{GVF_GWM_DO_PS}, where it was shown that three active devices were required to successfully shield an object from an incident field. The devices were represented by multipolar point sources determined from Green's formula and Graf's addition theorem for Hankel functions (see \cite{GVF_GWM_DO_2011}).

The body of work by Guevara Vasquez {\it et al.} inspired the recent papers by O'Neill {\it et al.} \cite{JO_OS_RCM_ABM_NVM,JO_OS_RCM_ABM_NVMCHM}, investigating an analytical model for Kirchhoff plates to cloak flexural waves scattered by a circular clamped void \cite{JO_OS_RCM_ABM_NVM} and a circular coated inclusion at resonant frequencies \cite{JO_OS_RCM_ABM_NVMCHM}. The method outlined in these papers is generic and relies on cancelling selected multipole orders of the scattered field with an appropriate choice of active source amplitudes, calculated analytically via a linear algebraic system of equations. The active sources are represented by the Green's function for the biharmonic operator, which has the additional advantageous property of being non-singular at the source point. The procedure outlined in these papers produces extremely efficient cloaking with only a small number of sources (as little as six were sufficient in \cite{JO_OS_RCM_ABM_NVM}). Here we  draw a distinction between shielding and cloaking; in both cases the ultimate aim is to ensure that an observer, exterior to the cloak, receives no information about the object's presence. In shielding problems, a quiet region surrounding the object is created where the wave field practically vanishes and in this region the object is virtually invisible. 
By focusing solely on the region exterior to the cloak and object, effective cloaking can still be achieved by  ensuring that the total wave propagating away from the sources and object is sufficiently small. The  method adopted in this paper, from \cite{JO_OS_RCM_ABM_NVM} and \cite{JO_OS_RCM_ABM_NVMCHM}, is an example of cloaking as opposed to shielding. This active cloaking technique is analytical and does not have convergence issues within the active devices. 
Following in the footsteps of Guevara Vasquez {\it et al.} \cite{GVF_GWM_DO_2009a}--\cite{GVF_GWM_DO_PS} and Norris {\it et al.} \cite{ANN_FAA_WJP1}, \cite{ANN_FAA_WJP2}, Futhazar {\it et al.} \cite{GF_WJP_ANN} presented an inverse problem for finding source amplitudes required to cancel scattering from a finite region within a Kirchhoff plate, including a discussion of the errors associated with the order of truncation for specific numerical examples.

Previous work by O'Neill {\it et al.} \cite{JO_OS_RCM_ABM_NVM}, \cite{JO_OS_RCM_ABM_NVMCHM} has only dealt with active cloaking in Kirchhoff plates for a single continuous object. It is therefore a natural extension to consider the active cloaking of a discrete structure in the form of a finite cluster of scatterers. To the best of our knowledge, this is the first instance where a finite cluster is cloaked using active sources; a finite cluster is of interest as there can be multiple scattering, and highly directional anisotropy can be induced within the cluster and this is representative of real systems that might need to be cloaked. 
There has also been related extensive research employing discrete structures as the cloaking mechanism. For example, Colquitt {\it et al.} \cite{DJC_ISJ_NVM_ABM_MB_RCM} use a discrete system of rods and point masses to approximate a regularised continuum square-shaped cloak. Effective cloaking was demonstrated for solutions to the Helmholtz equation in the low frequency regime. In optics, the paper by Nicorovici {\it et al.} \cite{NAPN_RCM_LCB} uses passive techniques in the context of plasmonic metamaterials to cloak a finite cluster of cylinders. The authors derive a method for calculating the relative local density of states describing how a source interacts with a finite system of $N$ coated cylinders. They demonstrate that cloaking can be achieved exterior to a system of three cylinders by coating them with a material chosen to have properties ensuring plasmonic resonance occurs in the quasi-static limit; the cloaking method described in \cite{NAPN_RCM_LCB} is an example of passive cloaking in optics. 

Inspired by photonic and phononic crystals, there has been great recent interest in the study of platonic crystals, which use the Kirchhoff theory to model  flexural vibrations of a thin elastic plate patterned with a periodic array of scatterers. The simplest type of scatterer is a rigid pin, and many publications have featured various associated problems, producing a rich resource of literature on pinned plate systems. This motivated us to cloak a  discrete structure of pins in this article. 
Infinite periodic structured plates pinned at regular points were studied by Mace  (see, for example, \cite{Mace2}), with much of this earlier work reviewed by Mead \cite{Mead}; the assumption of clamped pins allows for an exact solution for a periodic square array using a Fourier series method, which has been adopted more recently by Antonakakis \& Craster in a collection of papers \cite{TARVC1}-\cite{TARVCSG3}. Dispersion surfaces and band diagrams are used to predict a variety of interesting wave phenomena, including shielding, lensing and negative refraction \cite{TARVCSG3}. A complementary approach using multipole methods has been employed by McPhedran, Movchan \& Movchan in a series of papers \cite{ABM_NVM_RCM}-\cite{RCM_ABM_NVM_MB_MJAS}. Multipole expansions, consisting of sums of cylindrical Bessel functions, are ideally suited to solve problems involving periodic arrays of finite-radius cylindrical holes. For zero displacement boundary conditions, in the limit as the radius of the voids vanishes, we recover the case of rigid pins. Dispersion surfaces and band diagrams (see for example, Movchan {\it et al.} \cite{ABM_NVM_RCM}, McPhedran {\it et al.} \cite{RCM_ABM_NVM}) are consistent with those obtained for the Fourier series approach. The Bloch-Floquet analysis of infinite periodic systems also provides insight in the form of approximate ranges of frequencies for resonances in finite clusters of these periodic arrays (see, for example, Antonakakis {\it et al.} \cite{TARVCSG1},  McPhedran {\it et al.} \cite{RCM_ABM_NVM_MB_MJAS}).

For arrays of infinite gratings, one may employ an elegant methodology for exploring Rayleigh-Bloch modes (Movchan {\it et al.} \cite{NVM_RCM_ABM_CGP}, Poulton {\it et al.} \cite{CGP_RCM_NVM_ABM}, Haslinger {\it et al.} \cite{SGH_NVM_ABM_RCM1}) to identify localisation regimes and wave-trapping behaviour for scattering problems. By tuning the frequency of the system, the gratings act like Fabry-P\'{e}rot plates where repeated reflections lead to constructive interference and localisation. The associated eigenvalue problem yields the Bloch modes linked with these resonance regimes for specified angles of incident plane waves. Haslinger {\it et al.} \cite{SGH_NVM_ABM_RCM2} give a table of optimised frequencies for a range of incident angles $0 \le \theta_i \le \pi/3$, which is relevant both to finite numbers of gratings, and to finite clusters like those studied here. This Green's function approach (also employed by Evans \& Porter \cite{DVE_RP}, Linton \& Martin \cite{CML_PAM},  Foldy \cite{LLF}) leads to rapid numerical solutions, determining propagating frequencies for pinned clusters, that we seek to cloak in this article.

The concepts of Dirac cone dispersion and Dirac points originate in topological insulators and have more recently been transferred into photonics  \cite{LL_JDJ_MS}. 
They are associated with adjacent bands, for which electrons obey the Schr\"{o}dinger equation, that meet at a single point. Typically connected with hexagonal and triangular geometries in systems governed by Maxwell's equations, and most notably associated with the electronic transport properties of graphene \cite{AHCN_FG_NMRP_KSN_AKG},  analogous Dirac and Dirac-like points have recently been observed in photonic and phononic crystals. Even more recently, similar properties have been demonstrated for platonic crystals by Smith {\it et al.} \cite{MJAS_RCM_MHM} and McPhedran {\it et al.}  \cite{RCM_ABM_NVM_MB_MJAS}.

The existence of Dirac cones and Dirac points are generally associated with the symmetries of the geometry of the system. When two perfect cones meet at a point, with linear dispersion, the cones are said to touch at a Dirac point. In the vicinity of a Dirac point, electrons propagate like waves in free space, unimpeded by the microstructure of the crystal. In platonic crystals, the analogous points generally possess a triple degeneracy, where the two Dirac-like cones are joined by another flat surface passing through what is known as a Dirac-like point. This is analogous to the photonic terminology adopted by Mei {\it et al.} \cite{JM_YW_CTC_ZQZ} where the existence of linear dispersions near the $\Gamma$ point of the reciprocal lattice for the square array is the result of ``accidental" degeneracy of a doubly degenerate mode 
and a single mode. Sometimes known as a ``perturbed" Dirac point, the accidental degeneracy does not arise purely from the lattice symmetry (as for a Dirac point) but from a perturbation of the physical parameters, for example varying the radius of the inclusions (see Antonakakis {\it et al.} \cite{TARVCSG3}).

The structure of the paper is as follows. Section \ref{prob_form} formulates the problem of flexural wave scattering by a finite cluster of rigid pins in a Kirchhoff plate in the presence of active control sources. In this section we write the solution describing the out-of-plane displacement in a thin plate constrained by the finite cluster of pins surrounded by active sources. In section \ref{cloaked_pins}, we derive the algebraic system of equations that determine the unknown active source amplitudes which are required to deliver efficient cloaking. A detailed discussion on the broadband nature of our cloaking method, with the emphasis on choosing specific values of the spectral parameter from relevant band diagrams and dispersion surfaces for a thin plate pinned in a doubly periodic fashion, is in Section \ref{broadband_cloaking}. Finally, in Section \ref{results}, we present illustrative examples of effective cloaking  for various spectral parameter values; for higher frequencies, where the cluster generates a more complicated scattering pattern, we clarify the transition from locating the active control sources on the grid of the cluster to a circle surrounding the cluster, and demonstrate the advantages of this latter configuration. Concluding remarks are drawn together in Section \ref{conc_rem}.

\section{Formulation of the problem and governing equations}

In this section, we first present the governing equations for flexural wave propagation in a Kirchhoff plate pinned at points forming a finite cluster surrounded by active control sources. The amplitudes of these sources are then found by solving a system of linear algebraic equations. The broadband nature of this cloaking method is also discussed in detail, with the aid of band diagrams and dispersion surfaces associated with a doubly periodic pinned plate.

\subsection{Problem setting}
\label{prob_form}

\begin{figure}[H]
\begin{center}
\includegraphics[width=12.6cm]{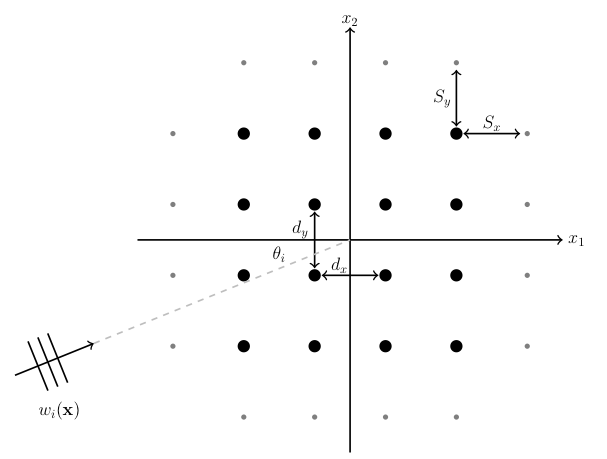}
\caption{Grid for a $4 \times 4$ $(N = 4)$ cluster of pins (larger black discs) surrounded by $K=4N =16$ active sources (smaller grey discs). A plane wave at an angle $\theta_i$ to the $x_1$-axis, $w_i ({\bf x})$, 
is incident upon the array. The spacings between the pins in the $x_1$ and $x_2$-directions are denoted by $d_x$, $d_y$ respectively. The spacings between the sources and the pins in the $x_1$ and $x_2$-directions are denoted by $S_x$ and  $S_y$, respectively. }
\label{config_4N_sources}
\end{center}
\end{figure}

Let a two-dimensional cluster of $N \times N$ pins, denoted by $\Pi$, be surrounded by $K$ active control sources (a particular example is shown in Fig.~\ref{config_4N_sources}). We define a pin as the limiting case of a circular void with a clamped boundary, that is, we require the displacement and its normal derivative to vanish, as the radius of the void tends to zero.  The origin of the axes coincides with the centre of the cluster, and a plane wave $W_i({\bf x};t)=W_i(x_1, x_2;t)=w_i({\bf x}) \exp{\{-i\omega t\}}$, where $\omega$ is the angular frequency, is incident at an angle $\theta_{i}$ to the $x_1$-axis. For the sake of simplicity, we consider a square array, such that the spacings $d_x, d_y$ between the pins in the $x_1$- and $x_2$-directions are equal. The spacings between the sources and the pins forming the boundary of the cluster, in the $x_1$- and $x_2$-directions, are denoted by $S_x$ and  $S_y$, respectively.

Let the locations of the pins be denoted by ${\bf x^p}=(x_1^{\,\bf p}, x_2^{\, \bf p})$, where ${\bf p}= (p_1,p_2)$ is a multi-index with integer components that describes the  square array $\Pi$,
and the position of the $m$-th active source  by ${\bf x}^{(s,m)}=(x_1^{(s,m)}, x_2^{(s,m)})$. Assuming time-harmonic vibrations, 
the out-of-plane displacement $w({\bf x})$ satisfies the following equation
\begin{equation}
\left(\Delta^2  - \beta^4 \right)w({\bf x}) + \sum_{{\bf p} \in \Pi} {\cal A}_{\bf p} \delta({\bf x} -{\bf x}^{\bf p}) +\sum_{m=1}^{K} {\cal B}_{m} \delta({\bf x} -{\bf x}^{(s,m)}) = 0, \quad {\bf x} \in {\Omega}, 
\label{opd_pins}
\end{equation}
where $\Delta^2$ is the biharmonic operator and $\beta$ is the spectral parameter with $\beta^2=\omega\sqrt{\rho h/D} \,$. Here $\rho, h$ and $D$ are the mass density, thickness and flexural rigidity of the plate $D=Eh^3/[12(1-\nu^2)]$, with $E$ being Young's modulus and $\nu$, Poisson's ratio of the elastic material.  The region ${\Omega}$ is exterior to the pins and sources,   ${\cal A}_{\bf p}$ represent the normalised transverse forces acting at the points ${\bf x^p}$, ${\cal B}_{m}$ represent the unknown complex amplitudes of the active sources and $\delta({\bf x} - {\bf x'})$ is the Dirac delta function centred at ${\bf x'}$. 
The solution to equation (\ref{opd_pins}) has the form
\begin{equation}
w({\bf x})=  \sum_{{\bf p} \in \Pi} {\cal A}_{\bf p} g({\bf x} -{\bf x}^{\bf p}) +\sum_{m=1}^{K} {\cal B}_{m} g({\bf x} -{\bf x}^{(s,m)}).
\label{opd_pins2}
\end{equation}
Here $g({\bf x} - {\bf x'})$ is the Green's function for the biharmonic operator which is non-singular at ${\bf x}={\bf x}'$ and is written as a sum of two Bessel functions
\begin{equation}
g({\bf x} - {\bf x'}) = \frac{i}{8 \beta^2} \left( H_0^{(1)}(\beta |{\bf x} - {\bf x'}|) +\frac{2i}{\pi}K_0(\beta |{\bf x} - {\bf x'}|)\right).
\label{Grn_fn_biharm}
\end{equation}

\subsection{Algebraic system for calculating active source amplitudes ${\cal B}_m$}
\label{cloaked_pins}

Given 
an incoming plane wave, $w_i (\bf x)$, 
incident on a finite cluster of rigid pins, we obtain the
following system of equations at the pin-position ${\bf x^q}$
\begin{equation}
w_i({\bf x^q}) + \sum_{{\bf p},{\bf q} \in \Pi} {\cal A}_{\bf p} g({\bf x^q}-{\bf x^p}) + \sum_{m=1}^{k} {\cal B}_m g({\bf x^q}-{\bf x}^{(s,m)}) =0,
\label{coeff_cond}
\end{equation}
where ${\bf p}$ and ${\bf q}$ are multi-indices in $\Pi$, the cluster of pins. Equation (\ref{coeff_cond}) in matrix form is
\begin{equation}
w_i({\bf x^q}) + G({\bf x^q}-{\bf x^p}){\textbf{A}} + G({\bf x^q}-{\bf x}^{(s,m)}){\textbf{B}} ={\bf 0}, \quad m=1,2,\dots,K,
\label{coeff_cond_matrix_form}
\end{equation}
where $G({\bf x^q}-{\bf x^p})$ and $G({\bf x^q}-{\bf x}^{(s,m)})$ are $N^2 \times N^2$ and  $N^2\times K$ matrices comprised of Green's functions for the biharmonic operator whose arguments depend on the distance between pin positions in the array $\Pi$, and the distance between pin positions and active source locations, respectively.
${\textbf{A}}$ and ${\textbf{B}}$ are  $N^2 \times 1$ and $K \times 1$  column vectors, respectively.

Using equation (\ref{coeff_cond_matrix_form}), we write the matrix ${\textbf{A}}$ of the coefficients ${\cal A}_{\bf p}$ in terms of the unknown source amplitudes ${\cal B}_m$ of the matrix $\textbf{B}$
\begin{equation}
\textbf{A} = -G^{-1}({\bf x^q}-{\bf x^p})w_i({\bf x^q}) - G^{-1}({\bf x^q}-{\bf x^p})G({\bf x^q}-{\bf x}^{(s,m)})\textbf{B}.
\label{Apin_cond}
\end{equation}
The amplitude of the total displacement $w_{tot}({\bf x}) $, at a point ${\bf x}$, exterior to the sources and array of pins, is given by
\begin{equation}
w_{tot}({\bf x}) = w_i({\bf x}) + G({\bf x}-{\bf x^p}) \textbf{A} + G({\bf x}-{\bf x}^{(s,m)})  \textbf{B},  \quad m=1,2,\dots,K.
\label{tot_field}
\end{equation}
In equation (\ref{tot_field}), $G({\bf x}-{\bf x^p}), \,G({\bf x}-{\bf x}^{(s,m)})$ are column vectors of size $N^2 \times 1$ and   $K \times 1$, respectively. For effective cloaking, we require the total field outside the region occupied by the pins and active sources to be approximately equal to the incident  plane wave
\begin{equation}
w_{tot}({\bf x}) \approx w_i({\bf x}). \label{tot_is_inc}
\end{equation}
The active source amplitudes ${\cal B}_m$ are to be found by cancelling out leading order terms of the scattered field and thereby ensuring that condition (\ref{tot_is_inc}) is satisfied. Eliminating the scattered field guarantees that an observer exterior to the active sources surrounding the finite array does not perceive the cluster's presence. Using condition (\ref{tot_is_inc}) together with equations  (\ref{Apin_cond}) and (\ref{tot_field}) we obtain
\begin{equation}\label{eqn_Bs}
\begin{split}
&w_i^{\top}({\bf x^q})G^{-1}({\bf x^q}-{\bf x^p})G({\bf x}-{\bf x^p}) \\
&\hspace{3.2cm}+ \left[ G^{\top}({\bf x^q}-{\bf x}^{(s,m)})G^{-1}({\bf x^q}-{\bf x^p})G({\bf x}- {\bf x^p}) - G({\bf x}-{\bf x}^{(s,m)})\right]\textbf{B} \approx 0,
\end{split}
\end{equation}
where we used the symmetry of the matrix $G({\bf x^q}-{\bf x^p})$.

We use Graf's addition theorem to re-expand entries of the matrices $G({\bf x}-{\bf x}^{(s,m)})$ and $G({\bf x}-{\bf x^p})$ about a general point ${\bf x}= R\, \text{exp}(i \theta), \, R>0$ in polar form.  For entries $g({\bf x}-{\bf x}^{(s,m)})$ of the matrix $G({\bf x}-{\bf x}^{(s,m)})$, we write the multipole representation
\begin{equation}\label{Gat_mat_source}
g({\bf x}-{\bf x}^{(s,m)})=  \frac{i}{8\beta^2} \sum_{l=-\infty}^{\infty} \left[ H_l^{(1)}(\beta R) J_l(\beta |{\bf x}^{(s,m)}|) + \frac{2i}{\pi} K_l(\beta R) I_l(\beta |{\bf x}^{(s,m)}|) \right] e^{i l (\theta - \text{arg}({\bf x}^{(s,m)}))},
\end{equation}
where ${\bf x}^{(s,m)} = |{\bf x}^{(s,m)}|\text{exp}\{i\, \text{arg}({\bf x}^{(s,m)})\}$ and $R>|{\bf x}^{(s,m)}|$. Similarly, entries of $G({\bf x}-{\bf x^p})$ can be written as equation (\ref{Gat_mat_source}) but with ${\bf x}^{(s,m)}$ replaced with ${\bf x^p}$.
 
Substitution of the propagating terms of equations (\ref{Gat_mat_source}) and their analogues for $g({\bf x}-{\bf x^p})$
into equation (\ref{eqn_Bs}) yields a system of linear algebraic equations for the unknown source amplitudes ${\cal B}_m,\, m=1,2,\dots,K$,  which eliminate the scattered field to a specific multipole order.  This order is connected with the number of active sources in the configuration set-up.
Thus, for the index $l$ running over the appropriate range, we obtain the following equations
\begin{equation}\label{alg_sys_eqns_B}
w_i^{\top}({\bf x^q})G^{-1}({\bf x^q}-{\bf x^p}){\bf J}_l( {\bf x^p}) + \left[ G^{\top}({\bf x^q}-{\bf x}^{(s,i)})G^{-1}({\bf x^q}-{\bf x^p}){\bf J}_l( {\bf x^p})  - {\bf J}_l( {\bf x}^{(s,m)}) \right] \textbf{B}= {\bf 0},
\end{equation}
where ${\bf J}_l( {\bf x^p})$ represents a matrix with entries $J_l(\beta {\bf |x^p|})\text{exp}\{-i l\, \text{arg}({\bf x^p})\}$ for all ${\bf p} \in \Pi$ and ${\bf J}_l({\bf x}^{(s,m)})$ represents a matrix with entries $J_l(\beta |{\bf x}^{(s,m)}|)\text{exp}\{-i l\, \text{arg}({\bf x}^{(s,m)})\}$ for all $m=1,2,\dots,K$. Note that ${\bf J}_l( {\bf x^p})$ and ${\bf J}_l({\bf x}^{(s,m)})$ are matrices of size $N^2\times 1$ and $K \times 1$, respectively.

\subsection{Selection of spectral parameter values to demonstrate broadband cloaking}
\label{broadband_cloaking}

\begin{figure}[h]
\begin{center}
\includegraphics[width=1\textwidth]{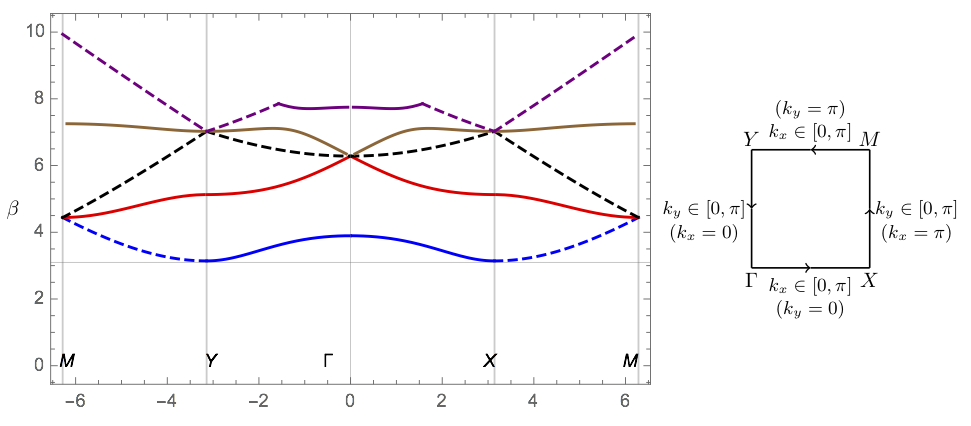}
\caption{Band diagram for square array of rigid pins with $d_x = d_y = 1$, using square Brillouin zone $\Gamma X M Y$ (the irreducible Brillouin zone is the triangle $\Gamma X M$ as in Movchan {\it et al.} \cite{ABM_NVM_RCM}).
The solid lines represent solutions of the dispersion relation (\ref{disp_rel1}), (\ref{disprel}), and
the dashed lines represent both dispersion curves and segments of the singularity lines (\ref{singular_lines}).
}
\label{bd_5bands}
\end{center}
\end{figure}

\begin{figure}[H]
\begin{center}
\includegraphics[width=1\textwidth]{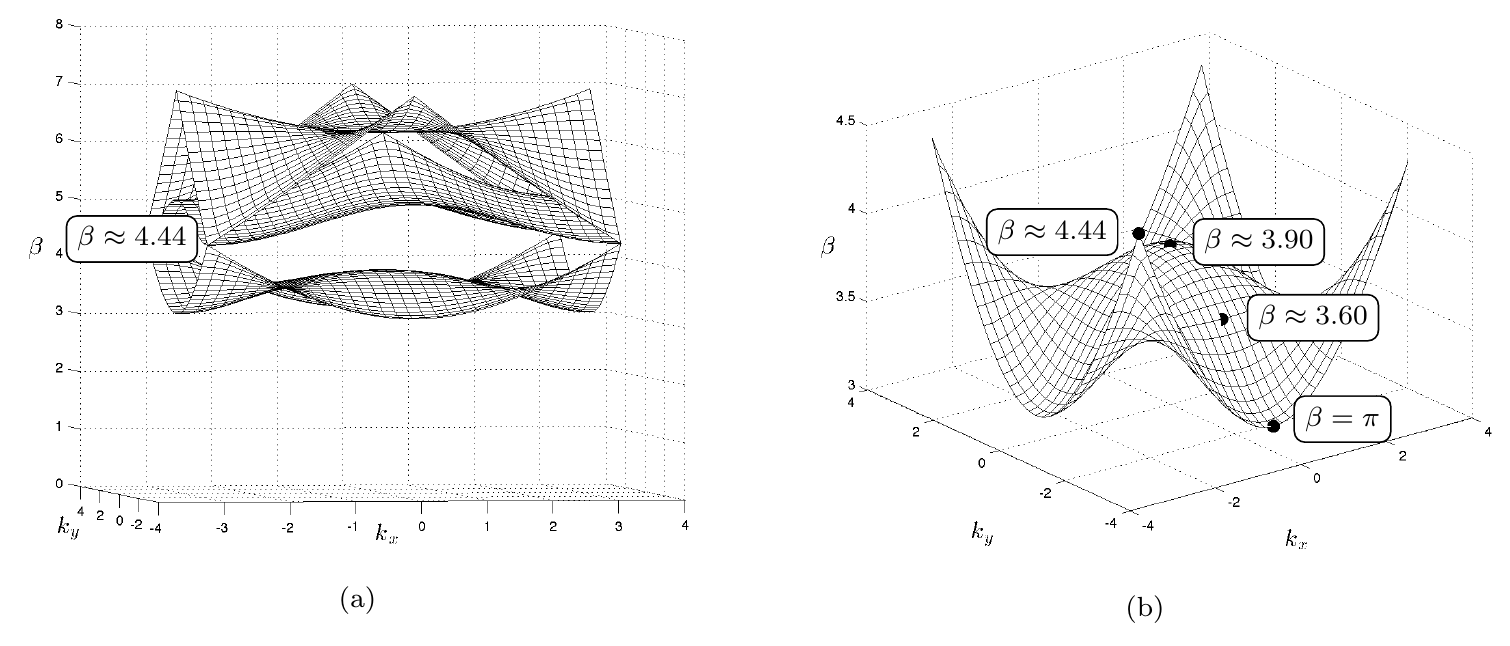}
\caption{(a) The first three dispersion surfaces shown together with the Dirac-like point highlighted. In (b) we display the first dispersion surface with the $\beta$-values of interest.}
\label{3_disp_surf}
\end{center}
\end{figure}

The method of cloaking employed here is a natural extension of that presented in the recent papers by O'Neill  {\it et al.} \cite{JO_OS_RCM_ABM_NVM}, \cite{JO_OS_RCM_ABM_NVMCHM}, where a single clamped void and a coated inclusion were cloaked using active sources, the former for non-resonant and the latter case for resonant regimes. As in these recent papers, the cloaking method implemented here for discrete structures is broadband, in the sense that we can produce efficient cloaking over a wide range of frequencies. We select illustrative frequencies from the band diagram (see Fig.~\ref{bd_5bands}) and the dispersion surfaces (see Fig.~\ref{3_disp_surf}) for a thin plate pinned in an infinite doubly periodic array. We also consider frequencies for examples of trapped flexural waves in finite arrays of gratings. Three typical regimes are identified from the dispersion surfaces:
\begin{itemize}
\item Zero frequency stop band and its edge,
\item Flat bands and standing wave frequencies,
\item Dirac-like points.
\end{itemize}

Figure \ref{bd_5bands} is the band diagram showing the first five dispersion curves for an infinite square array of rigid pins with $d_x=d_y=1$, 
first investigated by Movchan {\it et al.} \cite{ABM_NVM_RCM}. 

The dispersion curves plot the solutions of the dispersion relation for Bloch-Floquet waves in the infinite platonic crystal,
\begin{equation}\label{disp_rel1}
\frac{i}{8 \beta^2} \left( S_0^H (\beta, {\bf k}) + 1  + \frac{2i}{\pi} S_0^K (\beta, {\bf k}) \right)  = 0,
\end{equation}
expressed here in terms of lattice sums $S_0^H$ and $S_0^K$  (see, for example, Movchan {\it et al.} \cite{ABM_NVM_RCM} and McPhedran {\it et al.} \cite{RCM_ABM_NVM_MB_MJAS} for more details):
\begin{equation}
S_0^H  =  \sum_{q \ne (0,0)} H_0^{(1)} (\beta R_q) 
e^{i {\bf k} \cdot {\bf R}_q}\,; \quad \quad S_0^K = \sum_{q \ne (0,0)} K_0 (\beta R_q) 
e^{i {\bf k} \cdot {\bf R}_q}\,,
\label{disprel}
\end{equation}
where ${\bf k} = (k_x, k_y)$ denotes the Bloch wave vector and ${\bf R}_q = (R_q, \Theta_q)$ denotes the real-space array vector corresponding to the position of the pins in polar coordinates.
As first mentioned by McPhedran {\it et al.} \cite{RCM_ABM_NVM}, the dispersion bands for platonic crystals may be sandwiched between ``light lines" in reciprocal space (the dashed lines in Fig.~\ref{bd_5bands} represent segments of these light lines) which, for the square array, are defined by the equation
\begin{equation}\label{singular_lines}
\left(k_x + \frac{2 \pi h_x}{d_x} \right) ^2 + \left(k_y + \frac{2 \pi h_y}{d_y} \right) ^2  = \beta^2, \,\,\,\, h_x, h_y \in \mathbb{Z}.
\end{equation}
These singularity relations correspond to Bloch modes for the homogeneous 
plate, which for some frequency regimes, also satisfy the dispersion relation~(\ref{disprel}). Thus, the dashed lines in Fig.~\ref{bd_5bands} represent both dispersion curves and segments of the singularity lines. 

\begin{figure}[h]
\centering
\subfigure[]{
\includegraphics[clip,trim=1.7cm 1.1cm 1.7cm 1.1cm, width=0.31\textwidth]{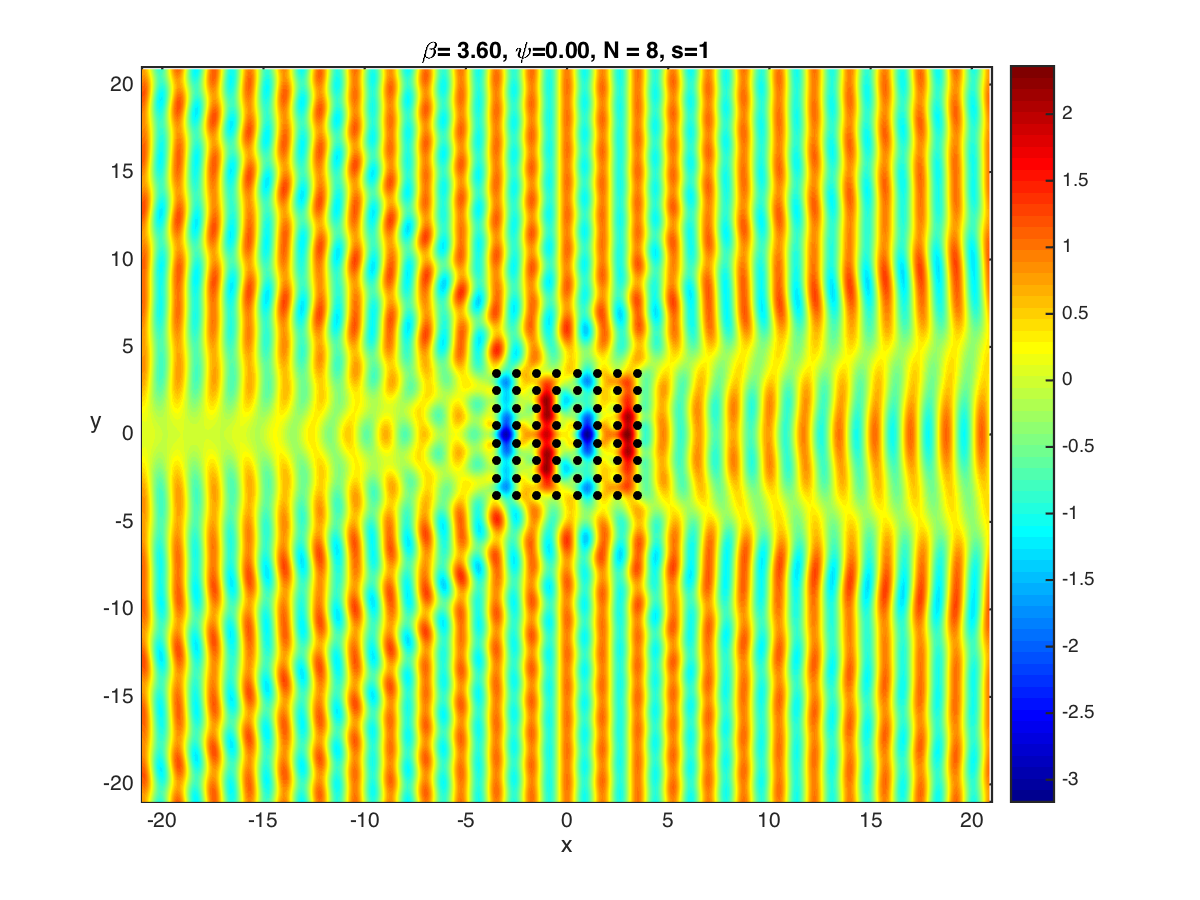}} ~~
\subfigure[]{\includegraphics[clip,trim=0.2cm 0.2cm 0.3cm 0.1cm,width=0.31\textwidth]{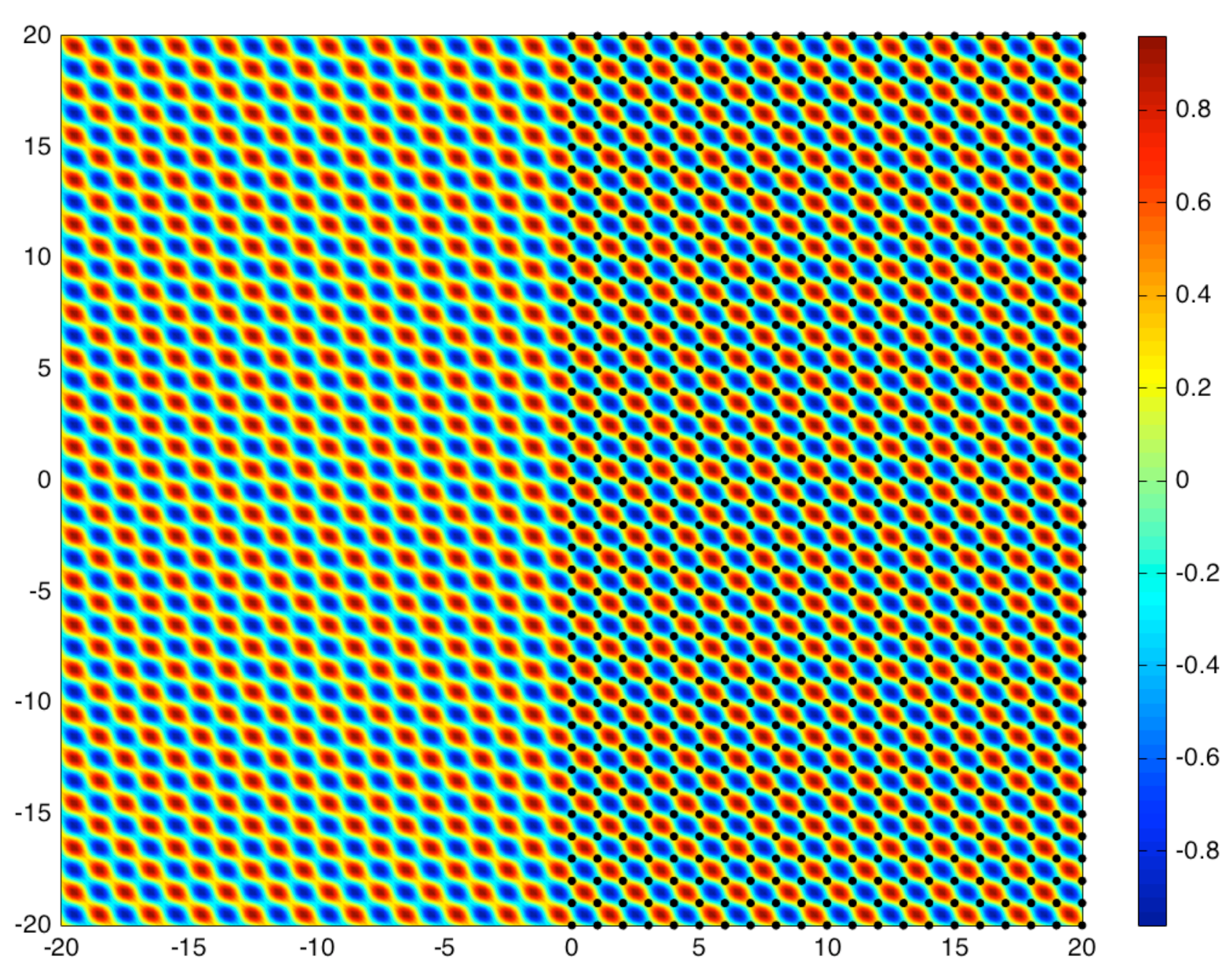}}
\subfigure[]{\includegraphics[clip,trim=0.0 0.2cm 0.3cm 0.1cm,width=0.32\textwidth]{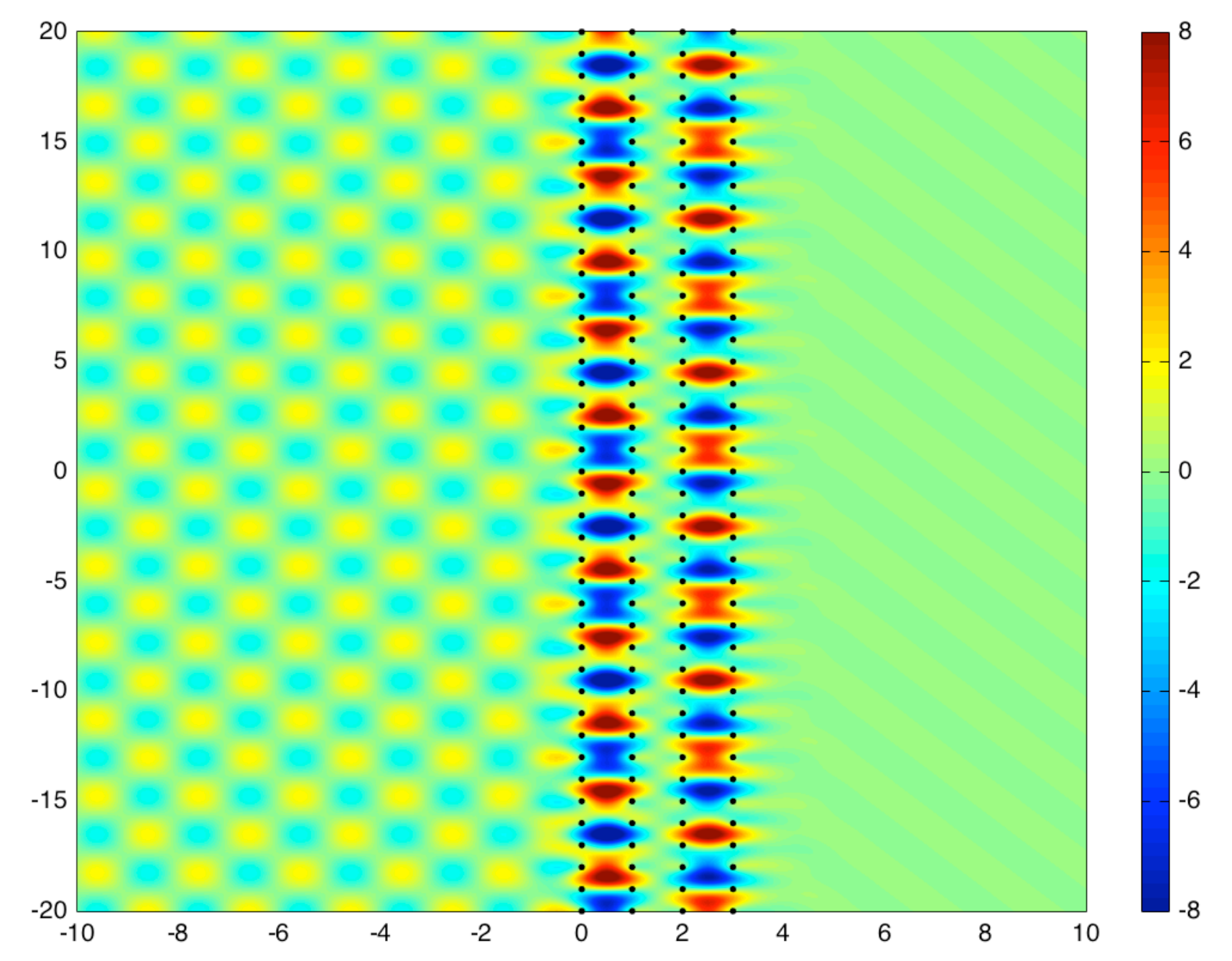}}
\caption{In all  figures here and in what follows, the real part of the amplitude of flexural displacement for the total field is shown, with its range, indicated by the colourbar, determined by the degree of localisation within the system. In addition, the pins are depicted by black discs and the horizontal and vertical axes are the $x_1$ and $x_2$-axes, respectively. (a) Plane wave normally incident on an uncloaked $8\times 8$ ($N=8$) cluster of pins. 
(b) Plane wave incident at $\theta_i = \pi/4$ on 2000 gratings with $d_x = d_y = 1$ for $\beta = 4.44$, $\kappa_y=3.1396$ corresponding to $M$ in the reciprocal lattice. (c) Localisation of flexural waves in a stack of 4 gratings with $d_x = d_ y =1$ for a plane wave incident at $\theta_i = \pi/6$ for $\beta = 3.60$. }
\label{neut_M}
\end{figure}

Stop bands (defined by $\beta < \pi$ for this square array, as shown in Fig.~\ref{bd_5bands}) correspond to frequency regimes for which waves do not propagate through the infinite periodic array of pins. For a finite cluster of pins at frequencies that fall within these ranges, the uncloaked pins scatter the waves, producing a shadow region behind the cluster. For frequencies in the pass bands ($\beta > \pi$ in this case as shown in Fig.~\ref{bd_5bands}), with a large enough cluster of pins, one would expect to see evidence of propagation through the pinned array. For an $8 \times 8$ cluster shown in Fig.~\ref{neut_M}(a), we see both propagation and scattering, which arises from an effective impedance mis-match between the array and the homogeneous plate exterior to the array. 
We use active sources to cloak the cluster and reconstruct the incident plane wave in the region exterior to the finite cluster and sources, thereby resembling the propagation properties of the infinite periodic system for pass band frequencies.

Flat bands (see the vicinities of $\Gamma$ for the first and fifth bands, and $X$ and $Y$ for the first and second, in Fig.~\ref{bd_5bands}) possess low group velocities $\partial \omega / \partial k \approx 0$, and are therefore associated with standing wave frequencies. The approach of high frequency homogenisation pioneered by Craster {\it et al.} (see, for example, \cite{RVC_JK_AVP}-\cite{RVC_JK_JP}, \cite{TARVC1}- \cite{TARVCSG1}) considers perturbations away from such standing wave frequencies, and is used to study dynamic wave propagation and associated problems. Standing wave frequencies and ``slow" waves are also synonymous with trapped modes and localisation, which are well known to arise from clamped boundary conditions (Poulton {\it et al.} \cite{CGP_RCM_NVM_ABM}).

As is typical for platonic crystals, the infinite square array of pins exhibits Dirac-like points, where a triple degeneracy arises from a flat surface passing through the intersection of two Dirac-like cones. 
Most Dirac and Dirac-like points are found on the boundary of the Brillouin zone, but can also be encountered at the centre of the Brillouin zone $\Gamma$ where ${\bf k} = (k_x, k_y) = (0, 0)$, provided that there is accidental degeneracy.
 An example of a Dirac-like point at $\Gamma$ is evident for $\beta = 2\pi$ in Fig.~\ref{bd_5bands}, where the third band (dashed curve) is the additional flat surface typical of pinned platonic crystal dispersion pictures. We also observe Dirac-like points on the edge of the Brillouin zone at $M$, $X$ and $Y$ in Fig.~\ref{bd_5bands}, as illustrated on the surface in Fig.~\ref{3_disp_surf}(a)  at $M$ for $\beta \approx 4.44$.  
Similar to the free propagation of electrons in the vicinity of Dirac points, there are directions of neutral propagation around a Dirac-like point in a platonic crystal, where the wave propagates with its direction unaffected by interaction with the microstructure. This effect of dynamic neutrality (which was considered as a form of cloaking by McPhedran {\it et al.}  \cite{RCM_ABM_NVM_MB_MJAS}) is also observed for subsets of the infinite periodic array. In Fig.~\ref{neut_M}(b), we show an example of a semi-infinite platonic crystal with a square array of pins similar to examples given by Haslinger {\it et al.}  \cite{SGH_RVC_ABM_NVM_ISJ}. The angle of incidence $\theta_i = \pi/4$ and the spectral parameter value of $\beta = 4.44$, ensures that we are in the neighbourhood of the Dirac-like point at $M$ in Figs.~\ref{bd_5bands} and in Fig.~\ref{3_disp_surf}(a). In Fig.~\ref{neut_M}(b) we observe the wave propagating through the array of pins without changing its direction and would expect to see a similar effect for a large enough finite cluster of pins in a square array. It is therefore very logical to analyse the improvement in cloaking upon the introduction of active sources.

We also select some $\beta$-values from known scattering and eigenvalue problems for subsets of the infinite doubly periodic array. Here we show an example for a stack of four gratings in Fig.~\ref{neut_M}(c); the choice of $ \theta_i = \pi/6$ and 
 $\beta = 3.60$, with the spacings $d_x = d_y = 1$, are the optimised parameter settings for localisation (see Table 1 in Haslinger {\it et al.} \cite{SGH_NVM_ABM_RCM2}). 
We observe high amplitudes of waves trapped within the gratings, which illustrate the Fabry-P\'{e}rot wave trapping capabilities of clusters of pins. 
In section \ref{cloaking_higher_freqs}, we demonstrate active cloaking at the angle of incidence and frequency for Fig.~\ref{neut_M}(c)
for a finite cluster of pins (the corresponding square subset of the finite stack of infinite gratings), where localisation effects are also observed.

\section{Results of active cloaking}
\label{results}

With our focus on achieving efficient cloaking, we begin by positioning the active control sources on the grid's rows and columns adjacent to the cluster of pins (see Fig.~\ref{config_4N_sources} with $S_x=S_y=d_x=d_y=1$)
and first set $K$, the number of sources, to be $4N$. 
Fig.~\ref{config_4N_sources} shows the general set up for a  a $4 \times 4\, (N=4)$ cluster of  pins surrounded by 16 sources. 
Naturally, we also consider the case of a $5 \times 5 \, (N=5)$ cluster of pins, hence distinguishing between  even and odd cases.
We then move on to position the active sources on a circle surrounding the finite cluster of pins, hence introducing flexibility in locating and choosing the number of these sources.

\subsection{Cloaking in the zero-frequency stop band}
\label{cloaking_stop_band}

The first $\beta$-value we select is from the zero-frequency stop band of the doubly periodic square array. 
We first consider positioning $K=4N$ sources exterior to the cluster, aligned with the rows and columns of the pins (see Fig.~\ref{config_4N_sources} for a schematic view of the configuration for the case when $N=4$). Secondly, we place four additional sources at the corners, creating a square array of sources that lies on the grid surrounding the cluster (hence $K=4(N+1)$). 
We plot the flexural displacement amplitudes of the total field for each configuration of sources, comparing them with  the uncloaked cases. 

\begin{figure}[H]
\begin{center}
\includegraphics[width=1\textwidth]{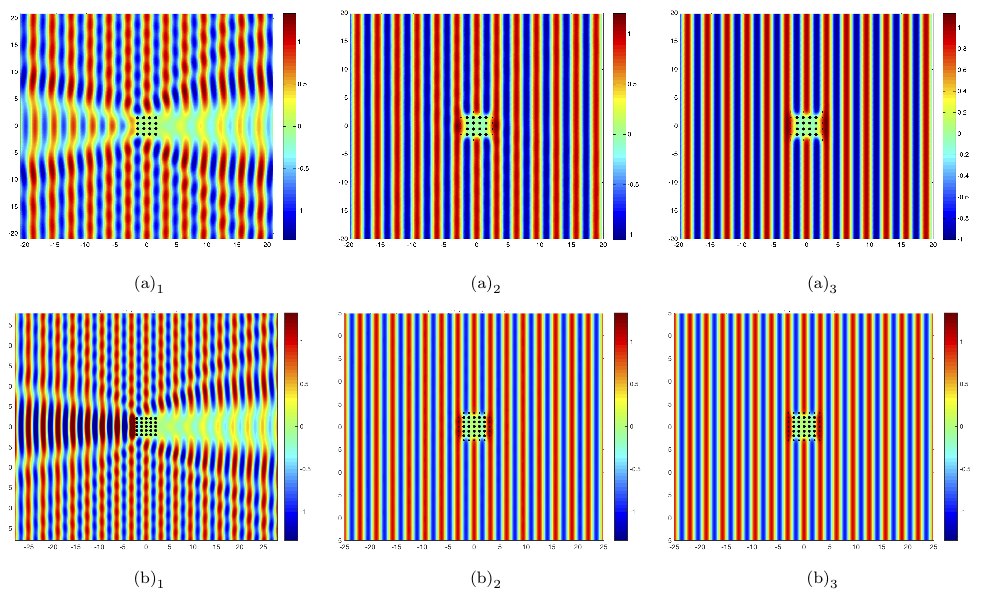}
\caption{Flexural wave scattering by (a) a $4\times 4$ ($N=4$) and (b) a $5\times 5$ ($N=5$) cluster of pins, depicted by large black discs. Plane wave is  incident at $\theta_i=0$ with $\beta =2.0$. Left: scattering with no active sources; middle:  cloaking with $K=4N$ active sources; right:  cloaking with $K=4(N+1)$ active sources. We note here and in what follows that the active source are illustrated by small black discs surrounding the cluster. 
}
\label{beta_2p0_N4_N5}
\end{center}
\end{figure}

Figure \ref{beta_2p0_N4_N5} shows the results of cloaking for $N=4$ (Fig.~\ref{beta_2p0_N4_N5}${\rm (a)}$) and $N=5$ (Fig.~\ref{beta_2p0_N4_N5}${\rm (b)}$) at $\beta=2.0$. In both cases, when no sources are present, waves do not penetrate the cluster and a large shadow region is created behind (see Figs.~\ref{beta_2p0_N4_N5}${\rm (a)}_1$ and  \ref{beta_2p0_N4_N5}${\rm (b)}_1$). 
Even for the small number of pins, the effect of the zero frequency stop band is seen in Fig.~\ref{beta_2p0_N4_N5}${\rm (a)}_1$ since virtually no waves pass through the cluster and there is strong reflection.
It should be noted here that clusters of 4 ($N=2$), or 9 ($N=3$), pins are too small to inherit such features from the infinite array and so are not considered in this paper. The choices of $N = 4$ and $5$ were made to represent the smallest finite arrays that inherit behaviour from the infinite array. Similarly small finite structures such as the clusters considered by Antonakakis {\it et al.} \cite{TARVCSG3}, and triplets of gratings by Haslinger {\it et al.}\cite{SGH_NVM_ABM_RCM2}, have previously been shown to exhibit properties of the corresponding infinite systems.  

In Figs.~\ref{beta_2p0_N4_N5}${\rm(a)}_2$ and \ref{beta_2p0_N4_N5}${\rm(b)}_2$, we see the reconstruction of the incident wave in the total field displacement amplitudes. There is no evidence of the shadow region and the field remains quiet inside the cluster. When sources are added along the diagonal at the corners, (see  Figs.~\ref{beta_2p0_N4_N5}${\rm(a)}_3$ and \ref{beta_2p0_N4_N5}${\rm(b)}_3$)
we see further improvement in cloaking for $N=4$, with wave fronts straightened compared to the no-corners case of Fig.~\ref{beta_2p0_N4_N5}${\rm{(a)}_2}$. For $N=5$, sources on the grid without corners is sufficient to produce near-perfect cloaking. 

\subsection{Cloaking at higher spectral values in the pass band}
\label{cloaking_higher_freqs}

We now extend our method to higher values of the spectral parameter $\beta$. Referring to Fig.~\ref{3_disp_surf}(b), we focus our attention to $\beta$-values of interest on the first dispersion surface, starting with $\beta=\pi$, the edge of the zero frequency stop band. We remark that the dispersion surfaces are obtained for an infinite plate pinned at a doubly periodic set of points and so they serve as a rough estimate for the $\beta$-values for which interesting features are observed for the finite cluster.

\begin{figure}[H]
\begin{center}
\includegraphics[width=1\textwidth]{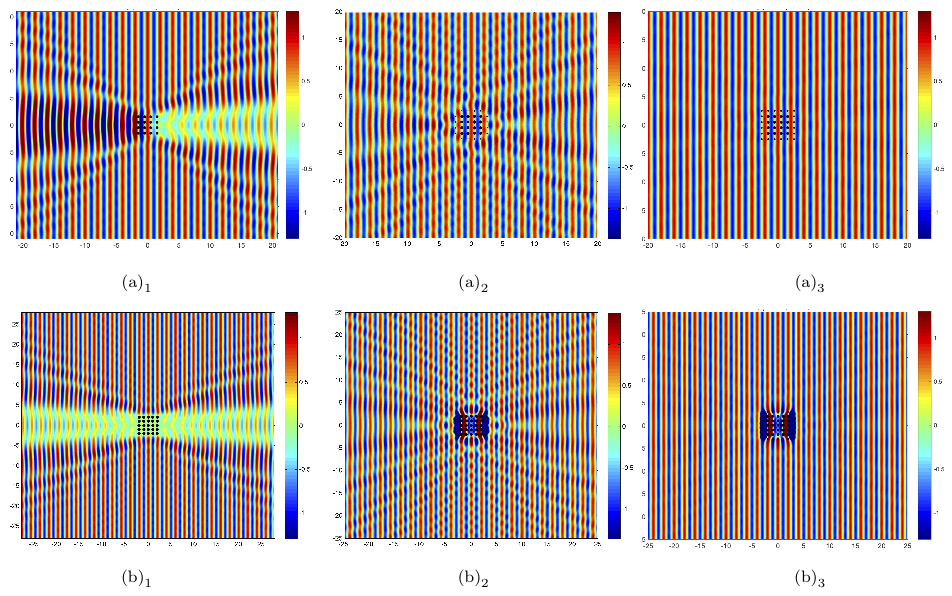}
\caption{Flexural wave scattering by a (a) $4\times 4$ ($N=4$) and (b) $5 \times 5$ ($N=5$) cluster of pins with an incident plane wave at an angle of $\theta_i=0$. From left to right: (left) no active sources, (middle) $K=4N$ active sources, (right) $K=4N+4$ active sources, all for $\beta=\pi$.}
\label{beta_eq_pi_all}
\end{center}
\end{figure}

Figures~\ref{beta_eq_pi_all}(a) and \ref{beta_eq_pi_all}(b) show amplitudes for a plane wave incident at $\theta_i=0$ with $\beta=\pi$ for the cases of $N=4$ and $N=5$, respectively.
Although $\beta = \pi$ lies on the boundary of the pass band for an infinite array, we see in Figs.~\ref{beta_eq_pi_all}${\rm (a)}_1$ and \ref{beta_eq_pi_all}${\rm (b)}_1$ that a large shadow region remains behind the finite cluster. 
It appears that the cluster is too small to inherit the exact frequency at which the pass band begins for the infinite structure. However, when active sources are introduced, the wave clearly penetrates through the finite cluster 
(see Figs.~\ref{beta_eq_pi_all}${\rm (a)}_{2,3}$ and \ref{beta_eq_pi_all}${\rm (b)}_{2,3}$). We observe that  adding sources at the corners of the diagonals of the cluster, illustrated in Figs.~\ref{beta_eq_pi_all}${\rm (a)}_3$ and \ref{beta_eq_pi_all}${\rm (b)}_3$, improves the cloaking further, and  the incident plane wave is successfully reconstructed in the total field exterior to the finite cluster and sources. 

We note that localised fields are encountered when cloaking the $N=5$ cluster, with amplitudes significantly greater than for the corresponding $N=4$ case. Here and in what follows, and only when necessary, we restrict the amplitude values to provide better comparison between corresponding cases and to improve the information gained about the cloaked fields outside the localised regions within the clusters.

\begin{figure}[H]
\begin{center}
\includegraphics[width=1\textwidth]{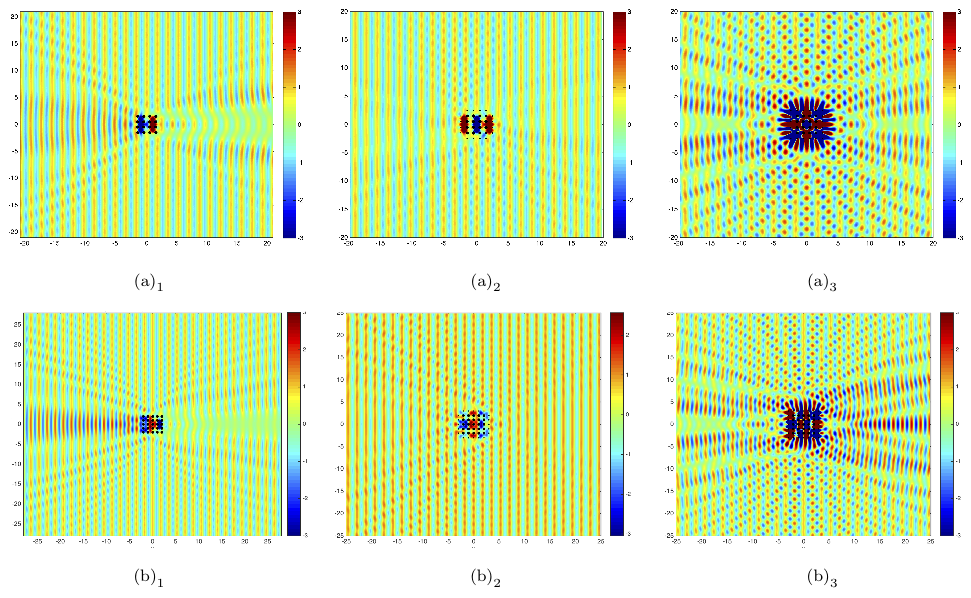}
\caption{Same as in Fig.~\ref{beta_eq_pi_all} but (a) shows the case of $N=4$ for $\beta=3.60$ and (b) shows the case of $N=5$ for $\beta=3.565$.}
\label{beta_3p60_cloaking_grid_breakdown}
\end{center}
\end{figure}

The next $\beta$-value we choose from the first dispersion surface in Fig.~\ref{3_disp_surf}(b) is $\beta \approx 3.60$, where the surface is locally flat at an inflexion. As flat sections of dispersion curves are associated with wave trapping and localisation, it is interesting to see how this affects the cloaking procedure. Since we are now attempting to cloak at a higher spectral parameter value, we note that efficient cloaking with sources  restricted to the grid ($K=4N$ or $K=4(N+1)$) may prove difficult.
We illustrate this potential hindrance in Fig.~\ref{beta_3p60_cloaking_grid_breakdown}.

Cloaking of the $4\times 4$ cluster when  $\beta=3.60$ indicates strong localisation effects, which is demonstrated in Fig.~\ref{beta_3p60_cloaking_grid_breakdown}(a). For the $5 \times 5$ cluster, we see strong localisation effects at a slightly lower value of $\beta = 3.565$, as shown in Fig.~\ref{beta_3p60_cloaking_grid_breakdown}(b). We emphasise that the dispersion surface in Fig.~\ref{3_disp_surf}(b) highlights key $\beta$-values for the infinite array and serves as a rough guide for us to pick spectral values of interest associated with the finite cluster. We also note that the variation in $\beta$ for which we see localisation  is due to internal resonances of the finite cluster.

In Figs.~\ref{beta_3p60_cloaking_grid_breakdown}${\rm (a)}_1$ and \ref{beta_3p60_cloaking_grid_breakdown}${\rm (b)}_1$, localisation dominates the field inside the cluster. 
Figures \ref{beta_3p60_cloaking_grid_breakdown}${\rm (a)}_{2,3}$ and \ref{beta_3p60_cloaking_grid_breakdown}${\rm (b)}_{2,3}$ display two unsuccessful cloaking attempts for each configuration. 
It is apparent that the cloaking efficacy is beginning to deteriorate at higher frequencies  if the sources are restricted to the grid, with or without additional sources located at the corners of the diagonals.

As a consequence of the above results, it is clear that more active sources are needed for efficient cloaking at higher frequencies. If we restrict the sources to the grid, this can be done either by moving the sources further away from the cluster, 
which, using our knowledge from our previous studies, is not so desirable or by adding more sources between those which are already present. 
We instead follow our recent work in active cloaking \cite{JO_OS_RCM_ABM_NVM}-\cite{JO_OS_RCM_ABM_NVMCHM}, and examine the effect of cloaking with active sources positioned on a circle surrounding the cluster, of radius $a_s$ centred at the origin. We ensure that two sources always lie on the $x_1$-axis, one in front and one behind the cluster, and the remaining sources are  distributed evenly, obeying  up-down symmetry.

\begin{figure}[t]
\centering
\subfigure[]{
\includegraphics[clip,trim=1.7cm 1.1cm 1.7cm 1.07cm,width=0.31\textwidth]{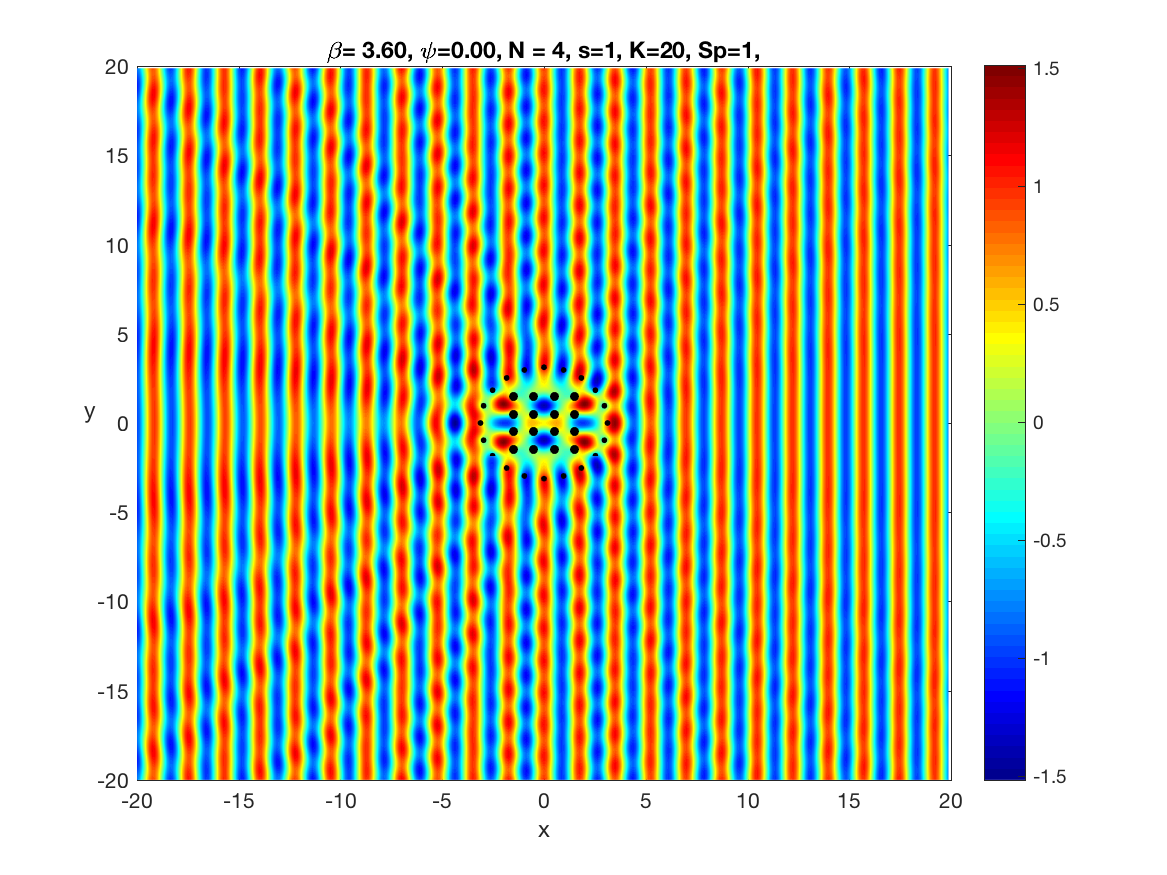}}~~
\subfigure[]{
\includegraphics[clip,trim=1.0cm 0cm 1.7cm 1.1cm,width=0.30\textwidth]{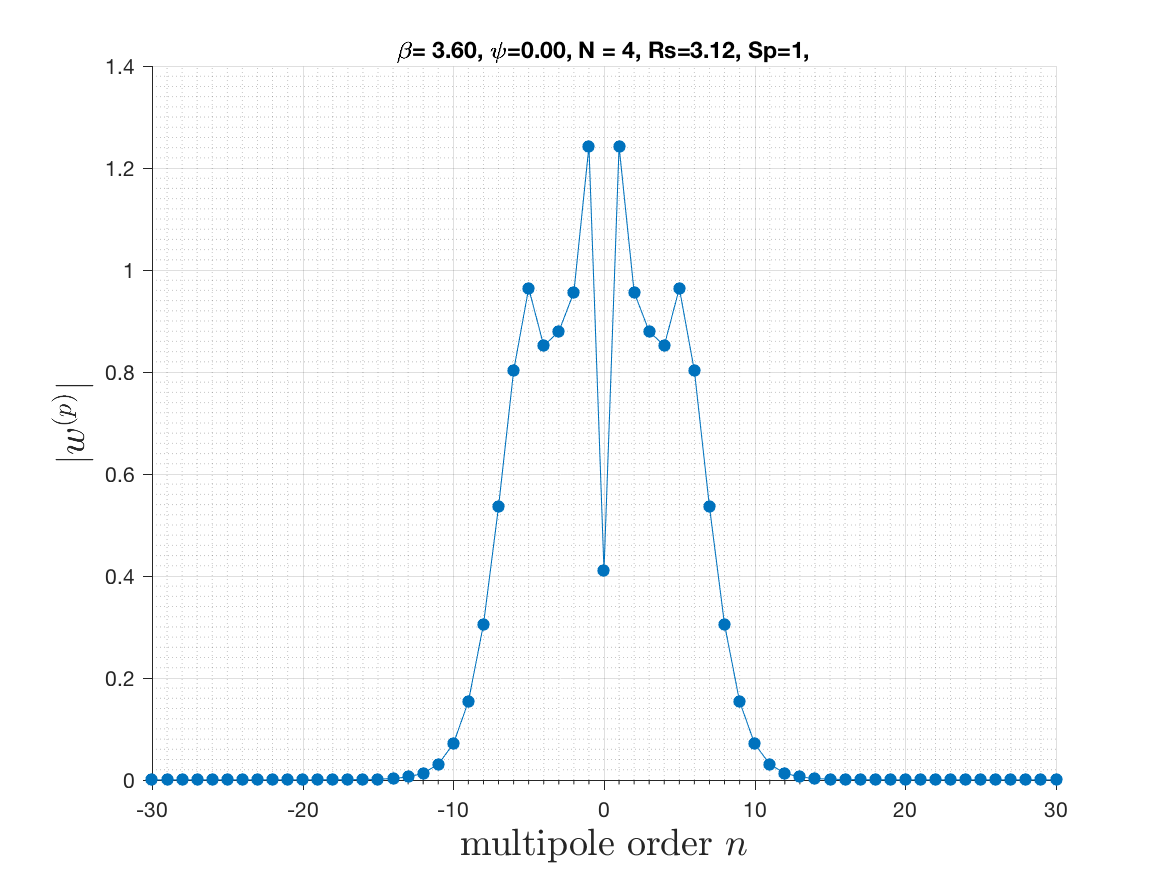}} ~~
\subfigure[]{
\includegraphics[clip,trim=1.7cm 1.1cm 1.7cm 1.1cm,width=0.31\textwidth]{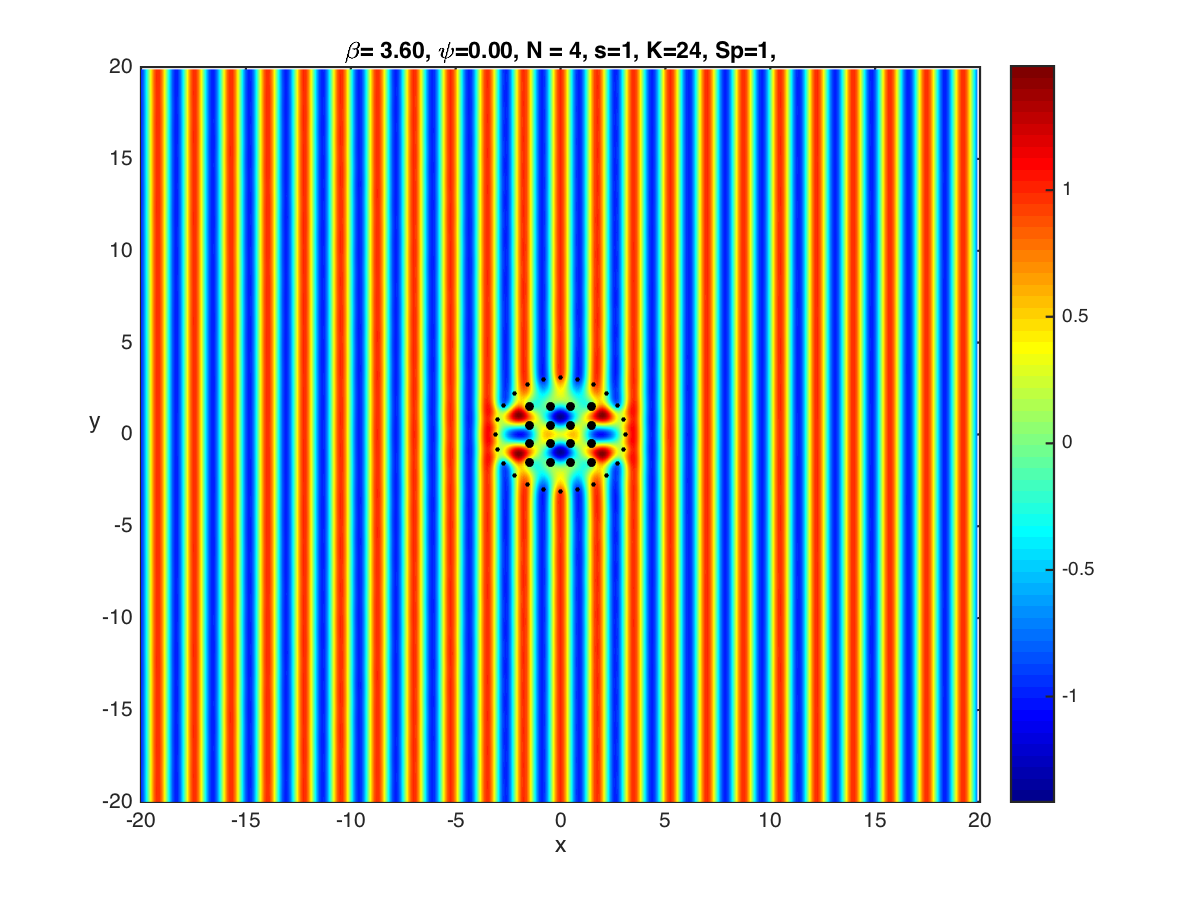}} 
\caption{For a $4\times 4$ ($N=4$) cluster of pins with an incident plane wave at an angle of $\theta_i=0$: (a) Flexural wave amplitudes for a cloaked cluster with $K=20$ evenly spaced sources on a circle of radius $a_s=(3\sqrt{2}+2)/2$, (b)  absolute value of the propagating part of the total field $|w^p|$ on the circle $a_s$ versus multipole order $n$, 
(c) flexural displacement amplitudes for a cloaked cluster with $K=26$, all for $\beta=3.60$.
}
\label{beta_3p60_circle}
\end{figure}

Retaining the maximum number of sources $K=20$ (see Fig.~\ref{beta_3p60_cloaking_grid_breakdown}${\rm (a)}_3$) used for a $4 \times 4$ cluster, but now locating them on a circle of radius $a_s=(3\sqrt{2}+2)/2$ (corresponding to a distance along the diagonal of 1 unit away from the corners of the cluster), we illustrate the resulting wave amplitudes  in Fig.~\ref{beta_3p60_circle}(a). This configuration brings an improvement in cloaking but the number of sources appears to be insufficient.

It is natural to consider how to estimate the number of sources required for effective cloaking. We address this by plotting the absolute value of the propagating part of the total field $w^p$ on the circle at which the sources are located, i.e. 
\begin{equation}\label{abs_propag_part}
|w^p| = \left|\sum_{{\bf p} \in \Pi} H_n^{(1)} (\beta a_s) J_n (\beta|{\bf x}^{{\bf p}}|) \exp{\{in \,\text{arg}({\bf x}^{{\bf p}})}\}\right|,
\end{equation}
 versus the multipole order $n$. This is shown in Fig.~\ref{beta_3p60_circle} (b), whence it can be deduced, by determining the multipole order  after which the contribution to the total field due to scattering is exponentially small, that the minimum number of sources required for efficient cloaking is $K=26$.  Finally, we plot amplitudes of the cloaked cluster using $26$ sources in Fig.~\ref{beta_3p60_circle} (c). 

\begin{figure}[H]
\centering
\subfigure[]{
\includegraphics[clip,trim=0.8cm 0cm 1.5cm 1cm,width=0.48\textwidth]{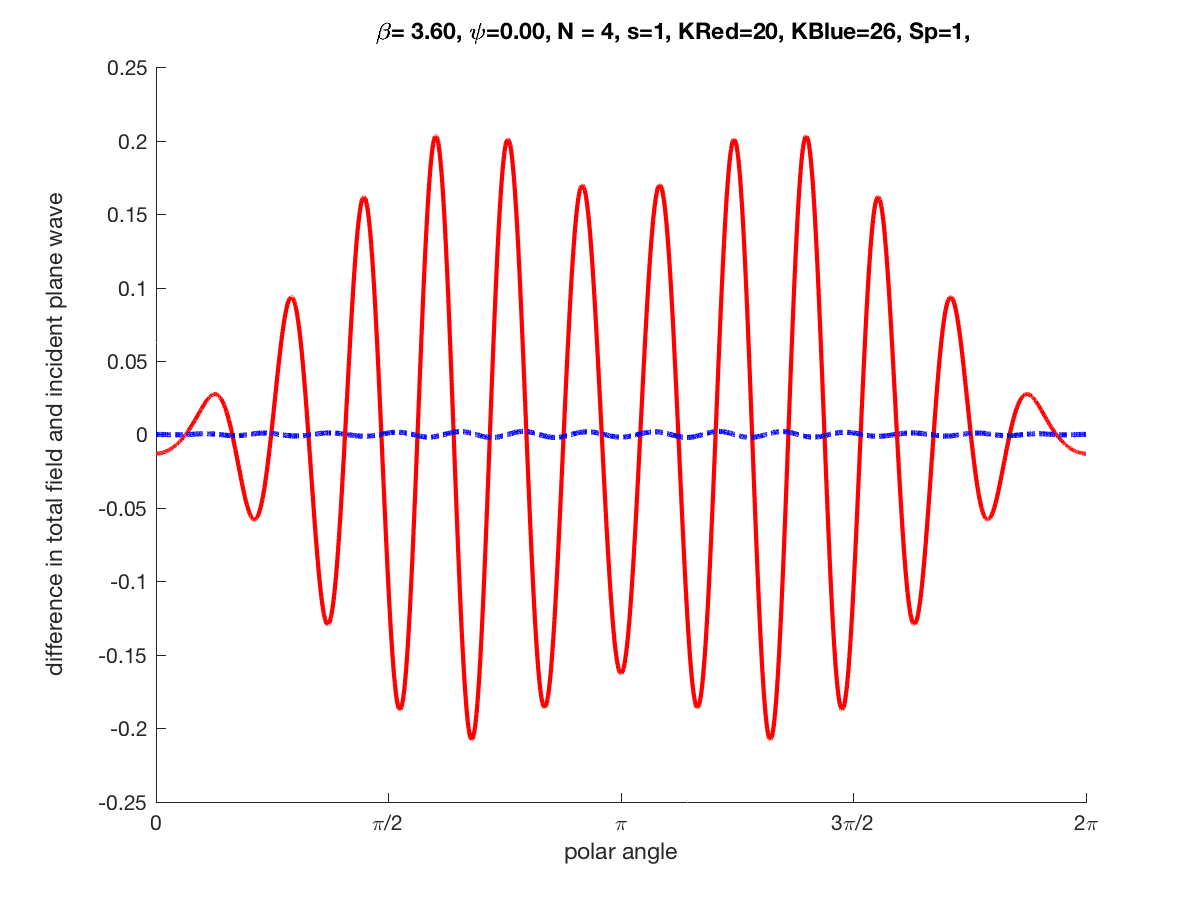}}~~
\subfigure[]{
\includegraphics[clip,trim=0.8cm 0cm 1.5cm 0.5cm,width=0.48\textwidth]{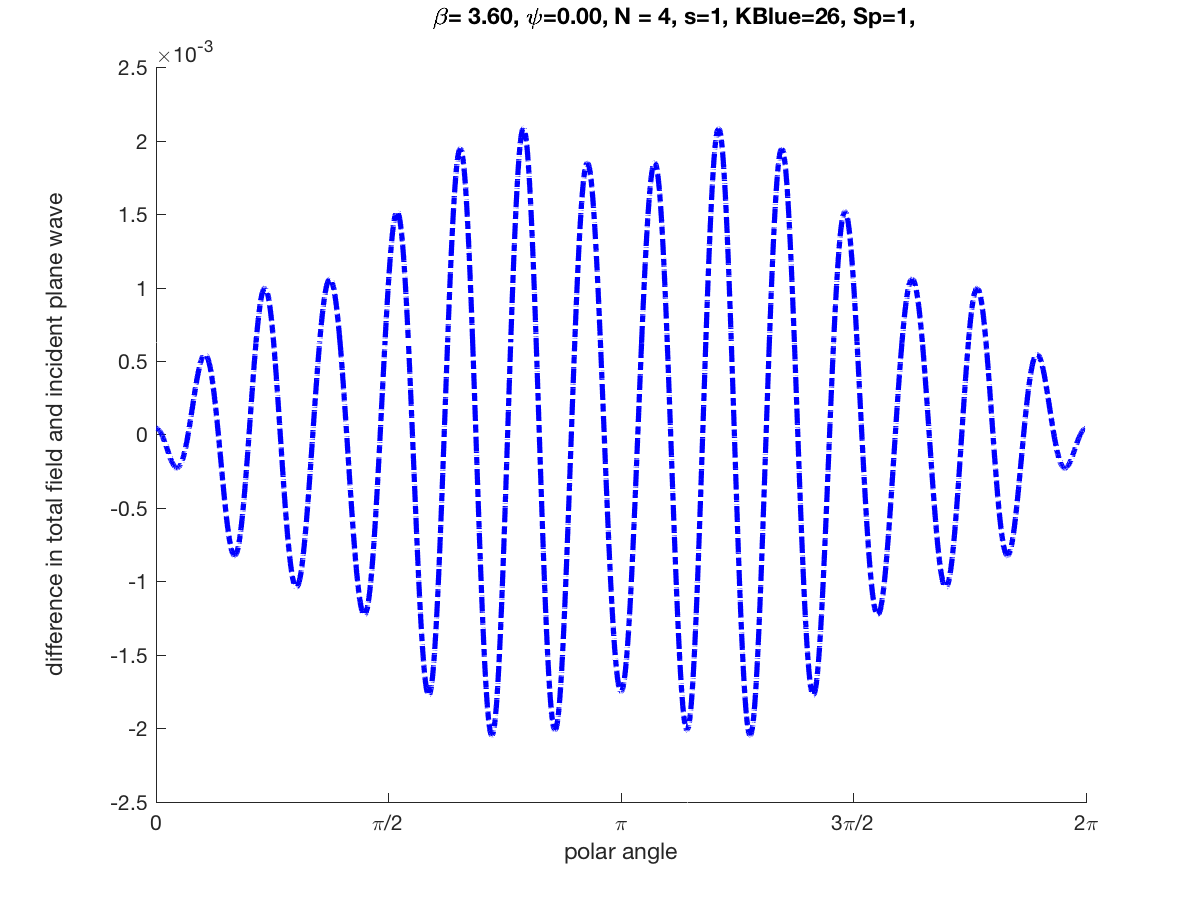}}
\caption{Difference in total wave amplitude and the incident plane wave evaluated on a circle of radius $15$ versus the polar angle. $4\times 4$ cluster of pins for $\theta_i=0$ and $\beta=3.60$. (a) $K=20$ (solid curve), $K=26$ (dashed curve) which is magnified in part (b) to clearly display its order.
}
\label{beta_3p60_totalfield_20_24}
\end{figure}

To further establish the efficiency of cloaking, we plot the difference between the total wave amplitude and the incoming plane wave evaluated at a circle of radius $15$ (sufficiently far away from the cluster and active sources) versus the polar angle. Fig.~\ref{beta_3p60_totalfield_20_24}(a) shows the comparison between 20 (solid curve) and 26 (dashed curve) sources located on a circle surrounding a $4 \times 4$ cluster for $\beta = 3.60$. It is clear that for the case of 26 sources, we see a significantly reduced difference between the two fields; the difference is of order $10^{-1}$ for $K=20$, but is of order $10^{-3}$ for $K=26$ (see  Fig.~\ref{beta_3p60_totalfield_20_24}(b)). 
In this article, we regard a cloaking to be satisfactorily efficient if the difference between the total wave amplitude and the incoming plane wave (absolute error) is of order $10^{-3}$ or less. We remark that the main body of the text from now on will include only the $N=4$ case and the analogous analysis for $N=5$ is placed in the Supplementary Material.

\begin{figure}[t]
\centering
\subfigure[]{
\includegraphics[clip,trim=1.8cm 1.1cm 1.7cm 1.07cm,width=0.48\textwidth]{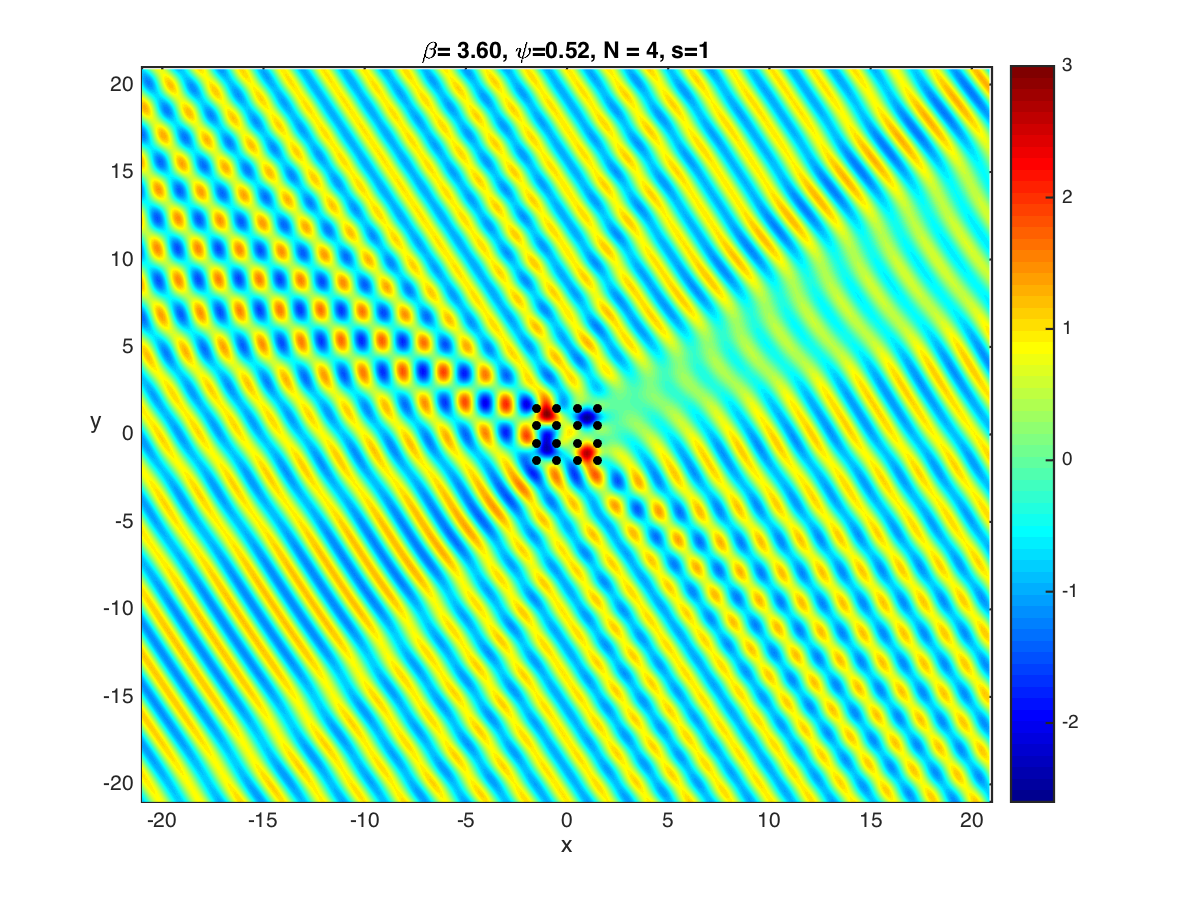}}~~
\subfigure[]{
\hspace{0.3cm}\includegraphics[clip,trim=1.8cm 1.1cm 1.7cm 1.07cm,width=0.48\textwidth]{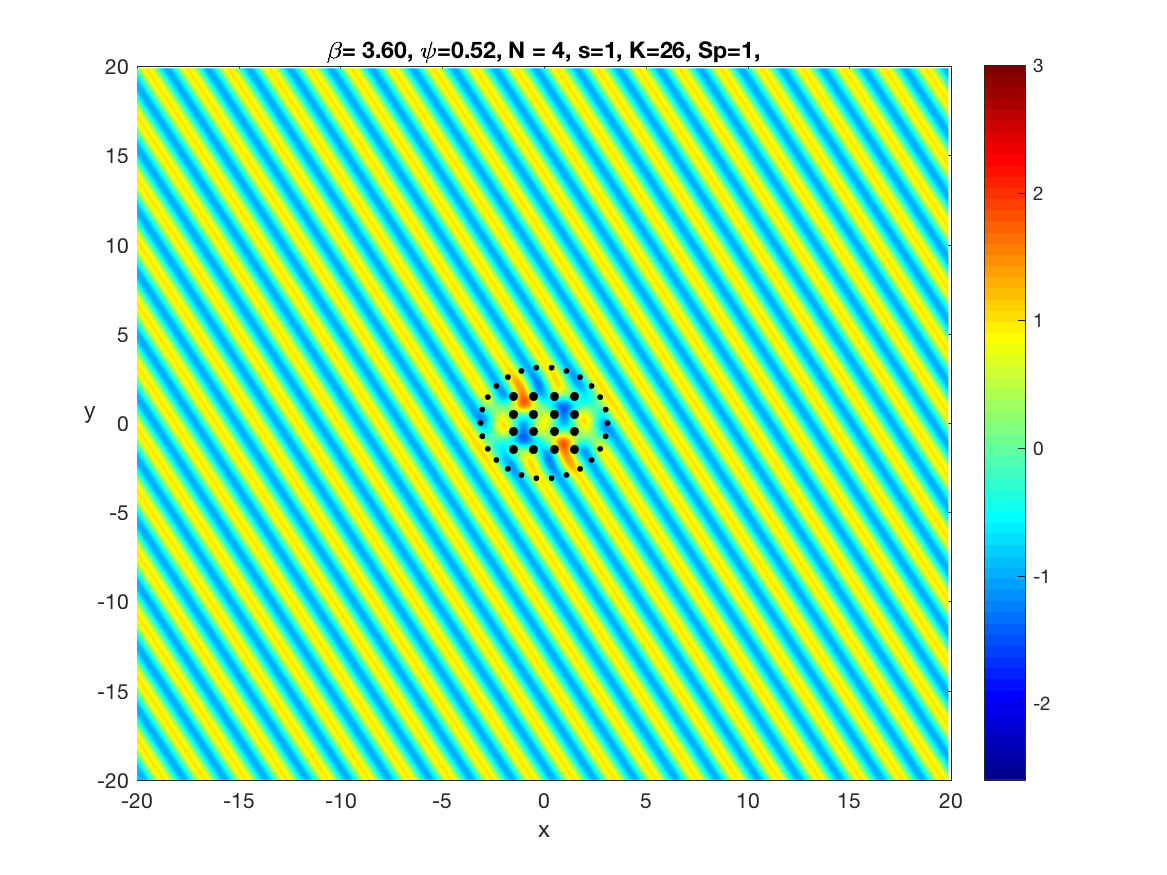}}\\
\hspace{-0.8cm}
\subfigure[]{
\includegraphics[clip,trim=1.0cm 0cm 1.7cm 1.1cm,width=0.445\textwidth]{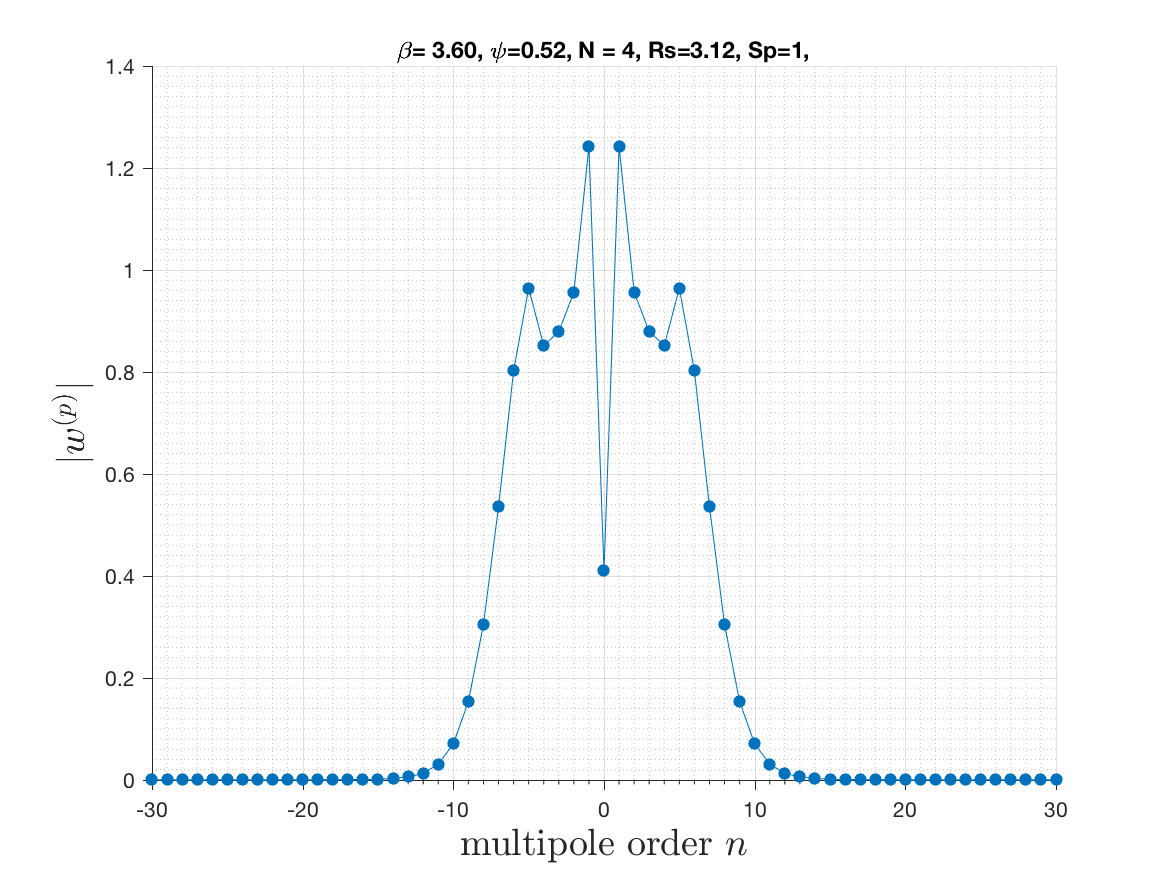}} ~~~~~~~
\subfigure[]{
\includegraphics[clip,trim=0.8cm 0cm 1.5cm 0.5cm,width=0.48\textwidth]{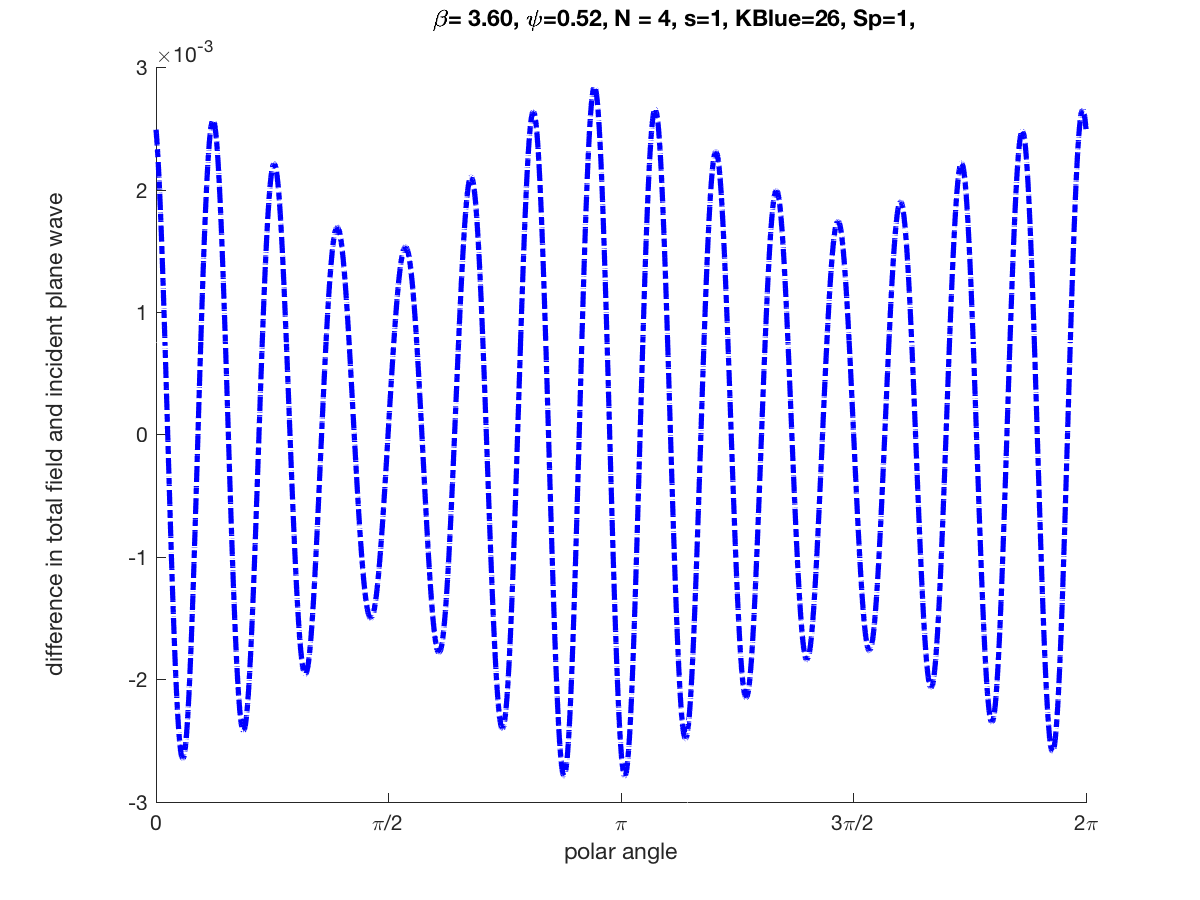}}
\caption{$4\times 4$ ($N=4$) cluster of pins with a plane wave incident at $\theta_i=\pi/6$ for $\beta=3.60$. Flexural wave amplitudes for (a) an uncloaked, (b) a cloaked cluster with $K=26$, (c) absolute value of the propagating part of the total field $|w^p|$ on the circle, of radius $a_s=(3\sqrt{2}+2)/2$, versus multipole order $n$, (d) absolute error evaluated on a circle of radius $15$ versus the polar angle.}
\label{beta_3p60_all_thetai_piby6}
\end{figure}

The method of cloaking outlined in section \ref{cloaked_pins} works equally well for oblique incidence. 
We illustrate this for $\theta_i=\pi/6$ and $\beta=3.60$ in Fig.~\ref{beta_3p60_all_thetai_piby6}, 
recalling that for a set of four gratings illustrated in Fig.~\ref{neut_M}(c), localisation was observed between the gratings. 
As expected, we see this localisation manifest itself as trapped waves in Fig.~\ref{beta_3p60_all_thetai_piby6}(a). However, this property is no hindrance to our cloaking method provided that we select the appropriate number of sources, as shown by Fig.~\ref{beta_3p60_all_thetai_piby6}(b). 
Using \ref{beta_3p60_all_thetai_piby6}(c), the minimum number of sources required for efficient cloaking of absolute error $10^{-3}$ is $K=26$. Note that this is the same number of sources required for the case of normal incidence  presented in Fig.~\ref{beta_3p60_circle}(b). 

\begin{figure}[H]
\centering
\subfigure[]{
\includegraphics[clip,trim=1.8cm 1.1cm 1.7cm 1.07cm,width=0.48\textwidth]{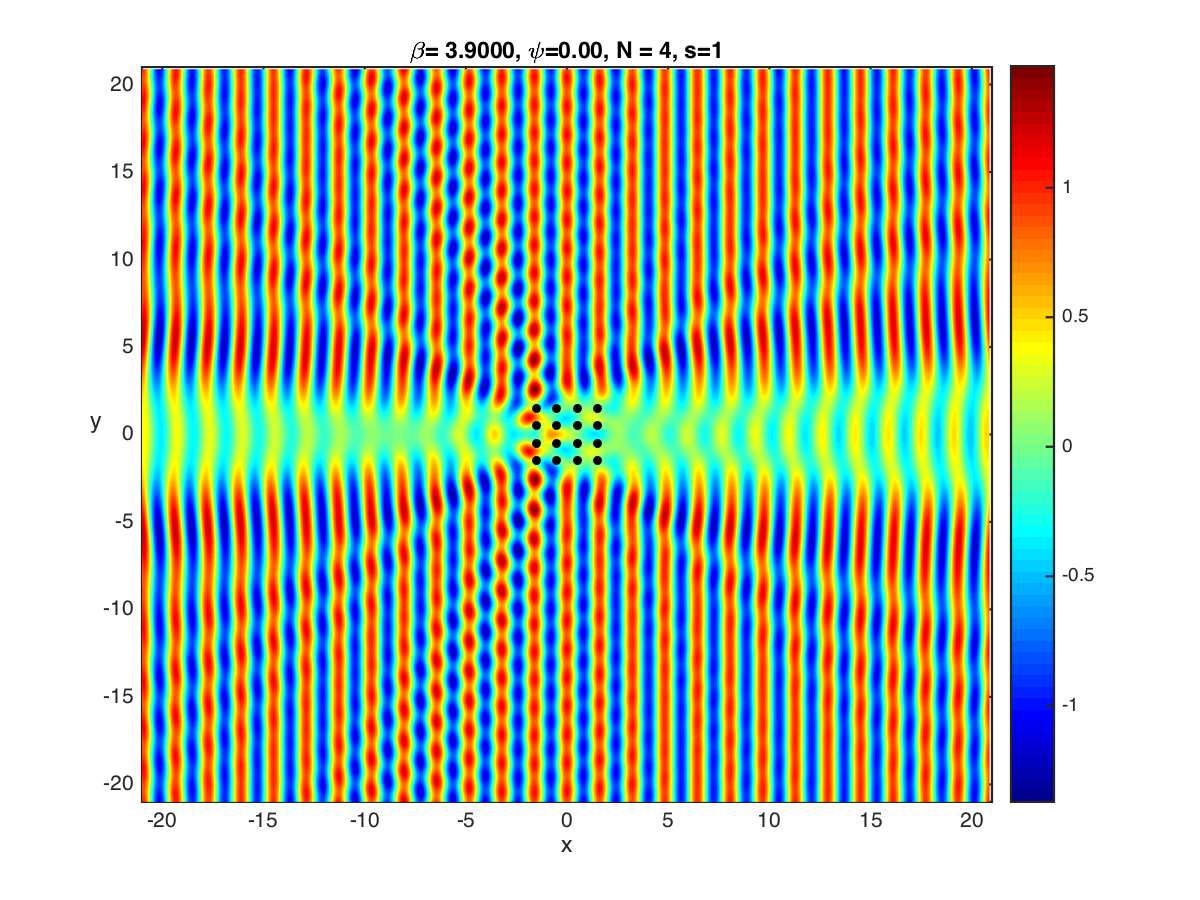}}~~
\subfigure[]{
\hspace{0.3cm}\includegraphics[clip,trim=1.8cm 1.1cm 1.7cm 1.07cm,width=0.48\textwidth]{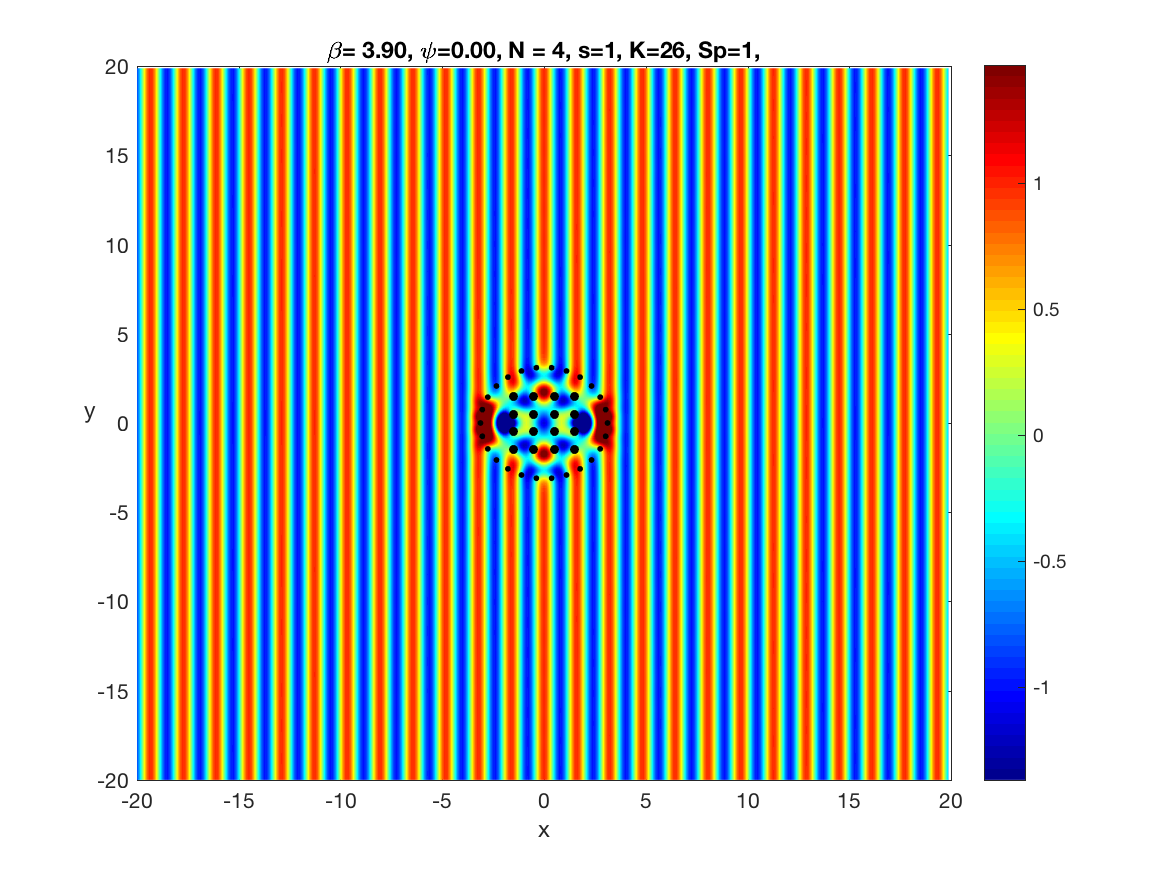}}\\
\hspace{-0.8cm}
\subfigure[]{
\includegraphics[clip,trim=1.0cm 0cm 1.7cm 1.1cm,width=0.445\textwidth]{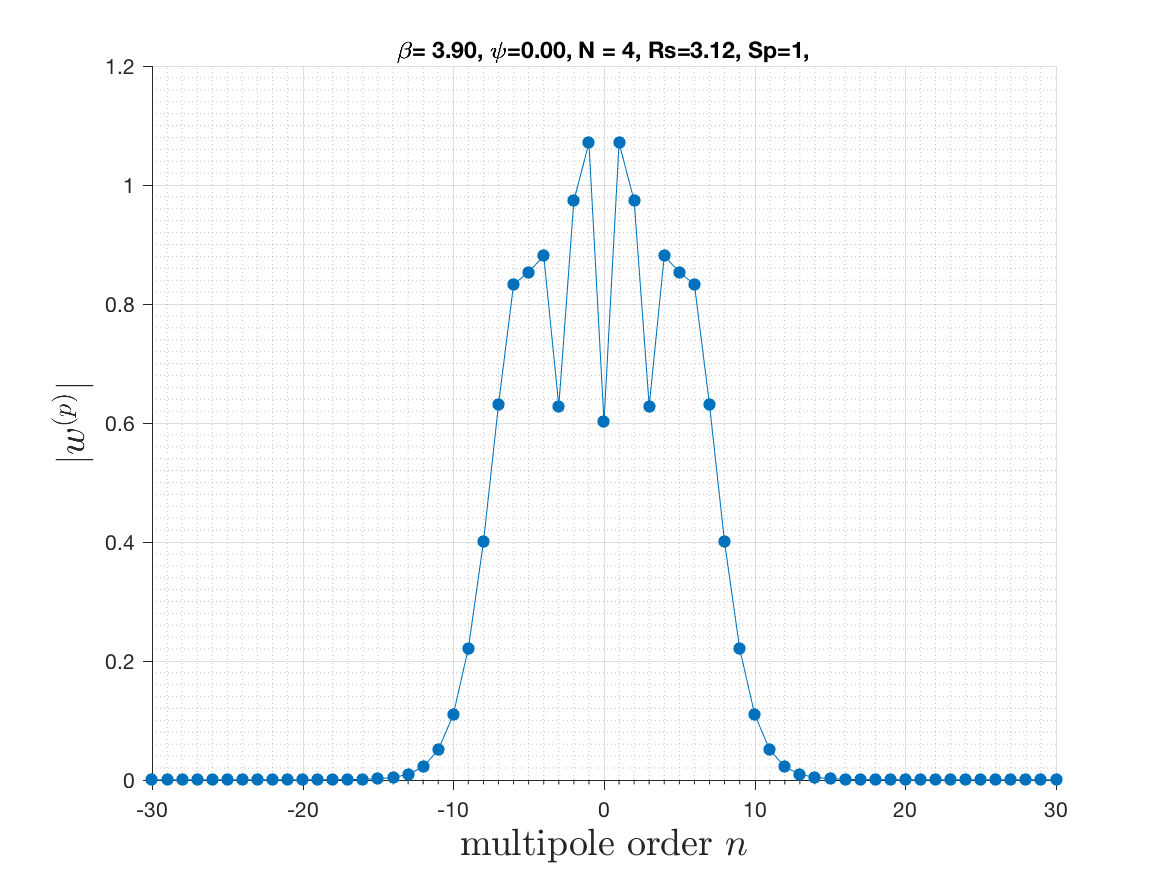}} ~~~~~~~
\subfigure[]{
\includegraphics[clip,trim=0.8cm 0cm 1.5cm 1.2cm,width=0.48\textwidth]{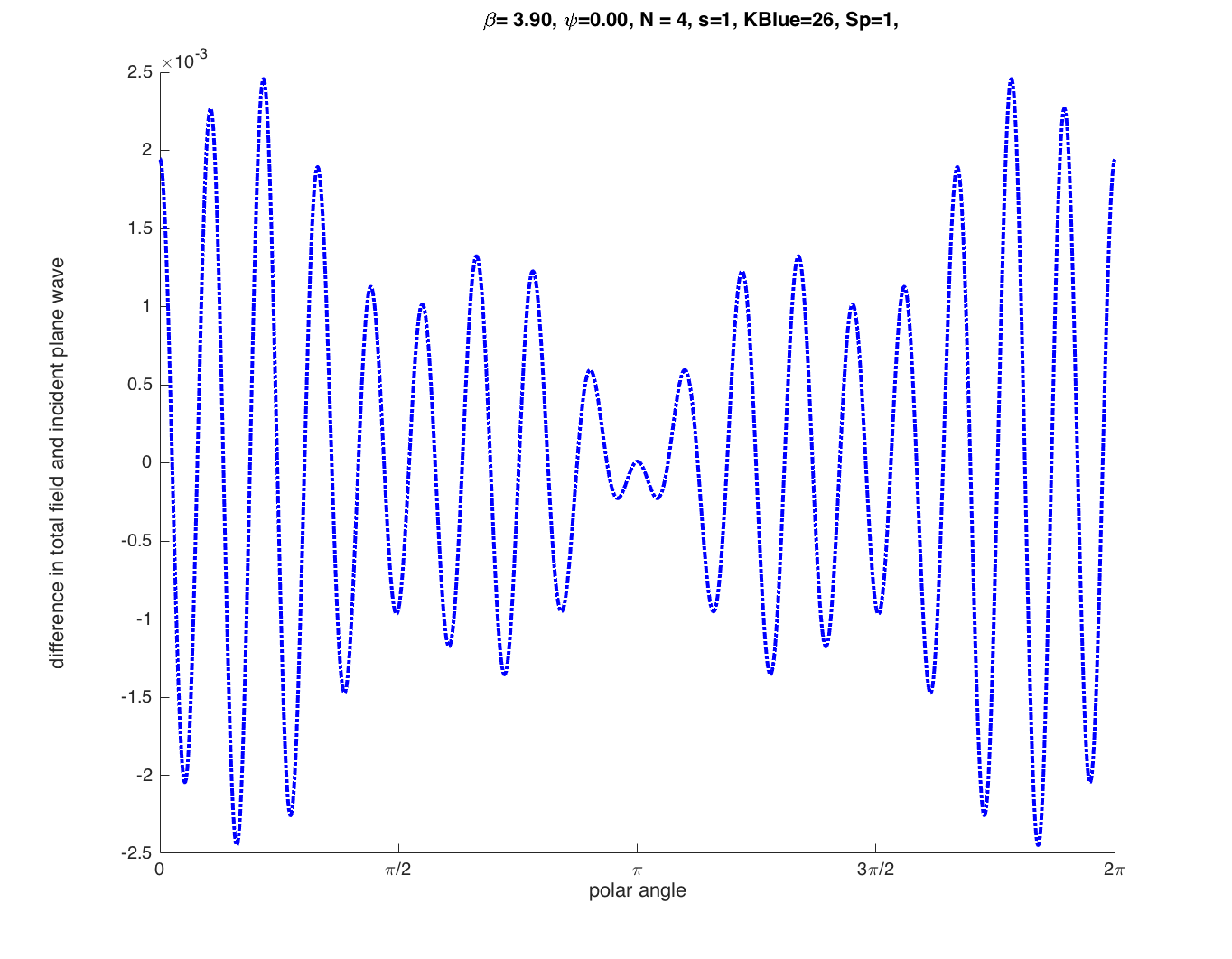}}
\caption{$4\times 4$ ($N=4$) cluster of pins with a plane wave incident at $\theta_i=0$ for $\beta=3.90$. Flexural wave amplitudes for (a) an uncloaked , (b) a cloaked cluster with $K=26$, (c) $|w^p|$ versus multipole order $n$ for a circle of radius $a_s=(3\sqrt{2}+2)/2$, (d) absolute error evaluated on a circle of radius $15$ versus the polar angle.}
\label{beta_3p90_all_thetai_0}
\end{figure}

\begin{figure}[H]
\centering
\subfigure[]{
\includegraphics[clip,trim=1.8cm 1.1cm 1.7cm 1.07cm,width=0.48\textwidth]{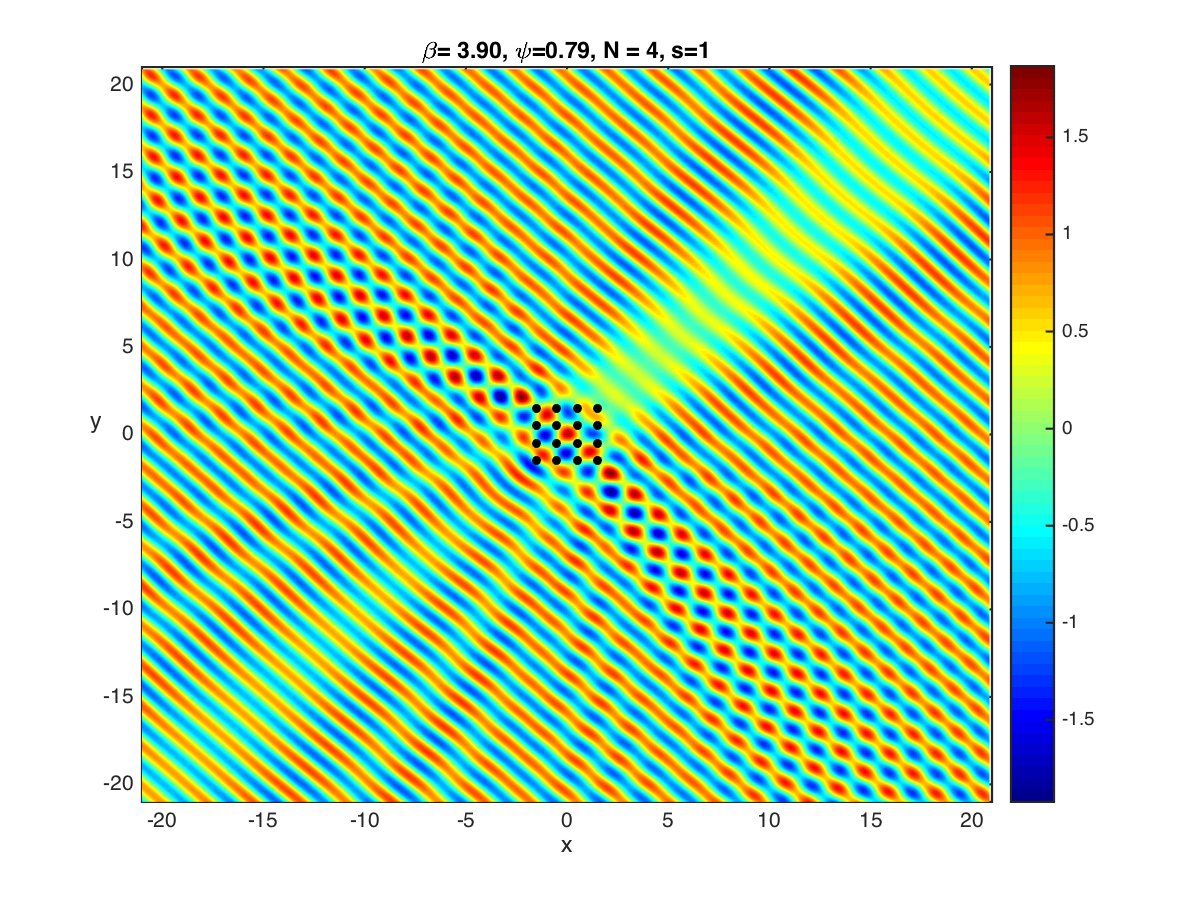}}~~
\subfigure[]{
\hspace{0.3cm}\includegraphics[clip,trim=1.8cm 1.1cm 1.7cm 1.07cm,width=0.48\textwidth]{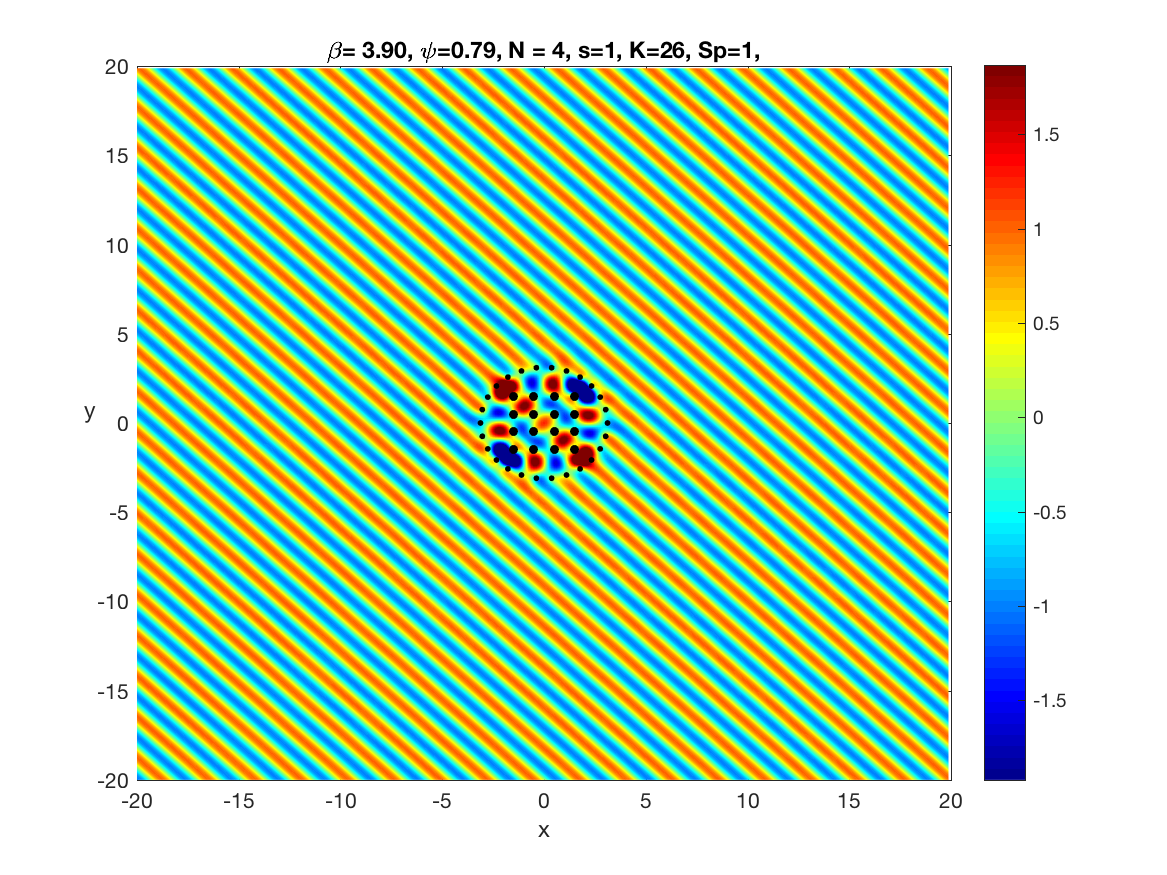}} 
\hspace{-0.8cm}
\subfigure[]{
\includegraphics[clip,trim=0.6cm 0cm 1.5cm 1.38cm,width=0.48\textwidth]{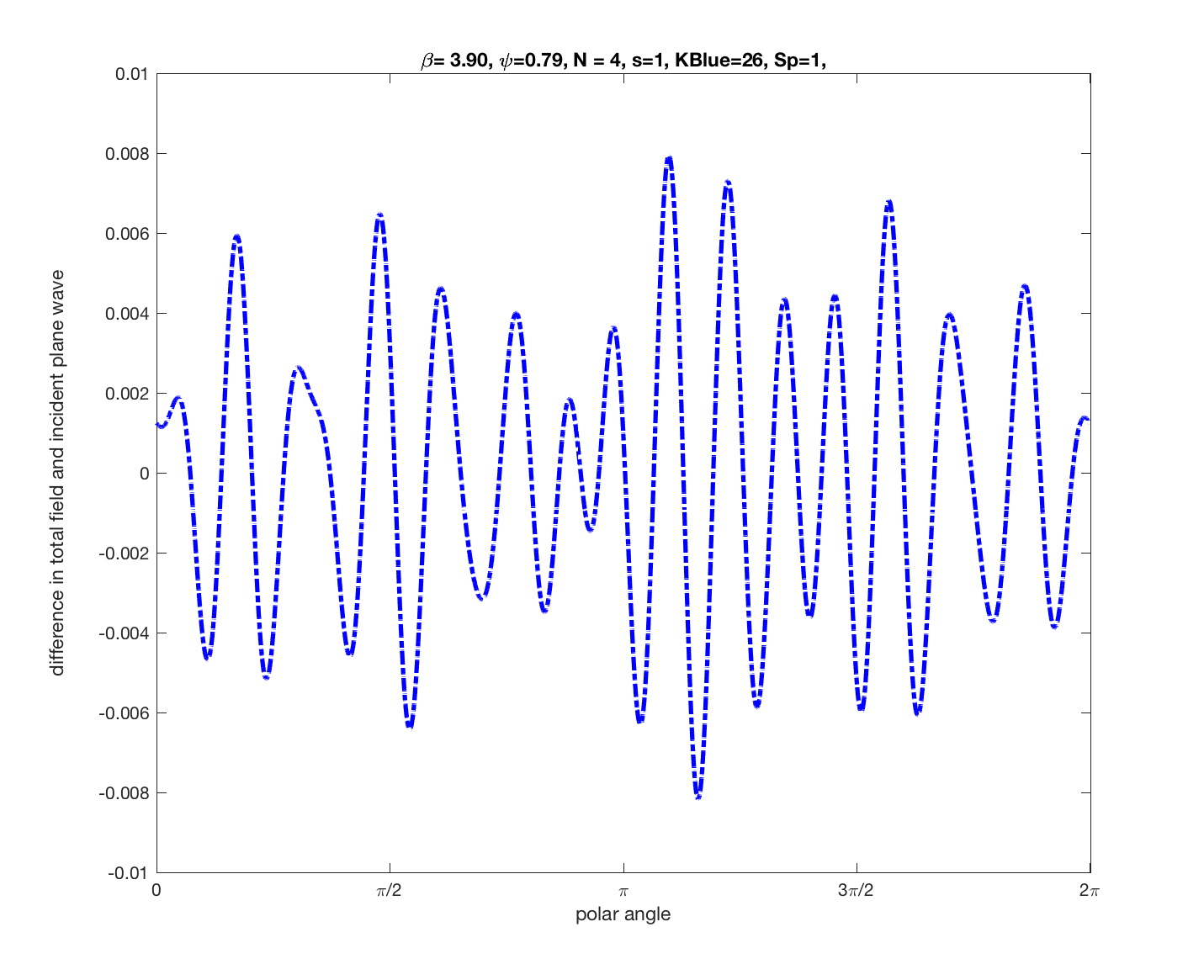}}
\caption{$4\times 4$ ($N=4$) cluster of pins with a plane wave incident at $\theta_i=\pi/4$ for $\beta=3.90$. Flexural wave amplitudes for (a) an uncloaked, (b) a cloaked cluster with $K=26$, 
(c) absolute error evaluated on a circle of radius $15$ versus the polar angle.}
\label{beta_3p90_all_thetai_piby4}
\end{figure}

Next, we turn our attention to cloaking at $\beta=3.90$, which is a local maximum and a standing wave frequency of the dispersion surface Fig.~\ref{3_disp_surf}(b). In Figs.~\ref{beta_3p90_all_thetai_0} and \ref{beta_3p90_all_thetai_piby4}, we demonstrate the results of our cloaking method for a plane wave incident at  angles of $\theta_i=0$ and $\theta_i=\pi/4$, respectively. In both cases, it is shown that 26 sources were required for effective cloaking.
We clearly reconstruct the incident plane wave, which is evident in both  plots of the total flexural displacement amplitudes (see Figs.~\ref{beta_3p90_all_thetai_0}(b) and \ref{beta_3p90_all_thetai_piby4}(b)). In parts (c) and (d) of Fig.~\ref{beta_3p90_all_thetai_0} and part (d) of \ref{beta_3p90_all_thetai_piby4}, the choice of the number of sources for effective cloaking is illustrated graphically for the desired accuracy. We note that $|w^p|$~(\ref{abs_propag_part}), plotted graphically in Fig.~\ref{beta_3p90_all_thetai_0}(c) versus multipole order for $\beta = 3.90$, is independent of $\theta_i$ and therefore we omit the equivalent plot for $\theta_i=\pi/4$ from Fig.~\ref{beta_3p90_all_thetai_piby4}. 

\begin{figure}[H]
\centering
\subfigure[]{
\includegraphics[clip,trim=1.8cm 1.1cm 1.7cm 1.07cm,width=0.48\textwidth]{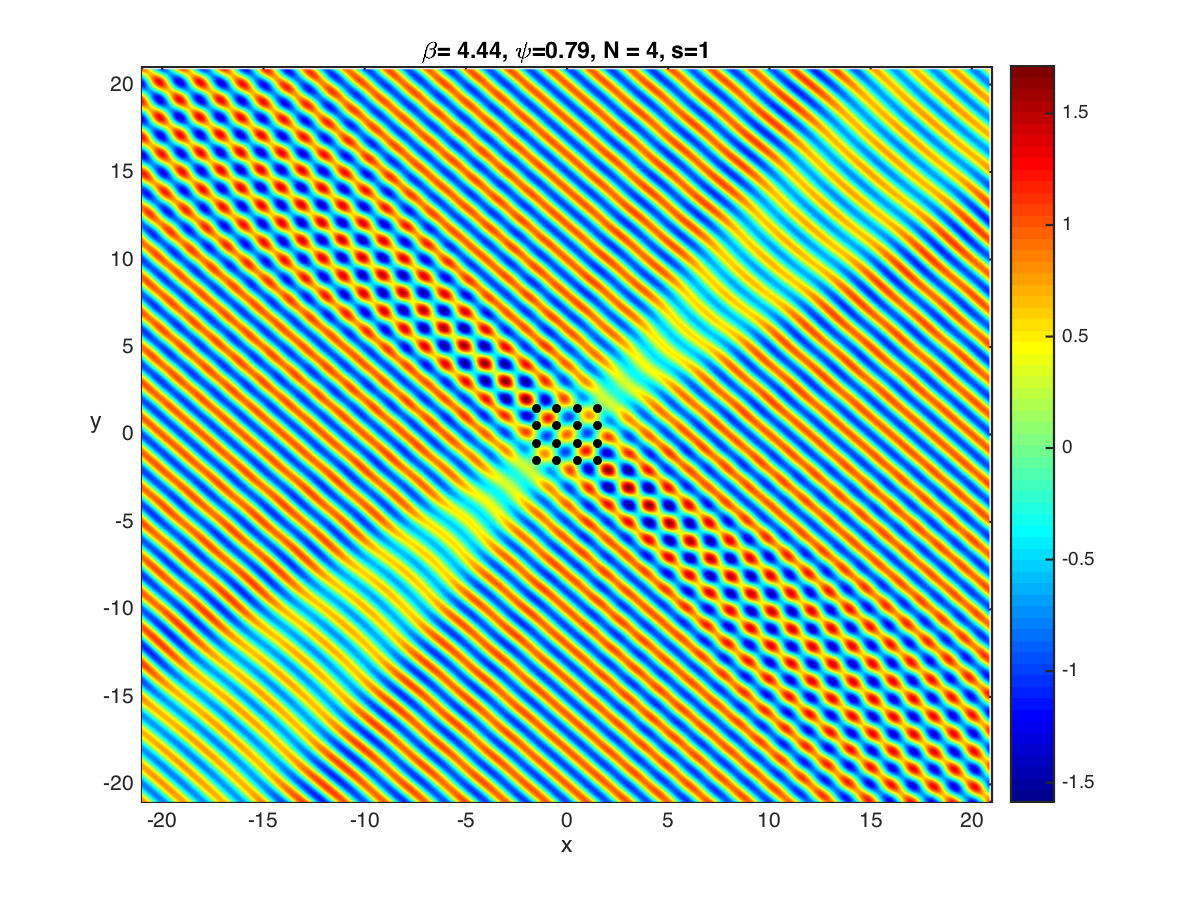}}~~
\subfigure[]{
\hspace{0.3cm}\includegraphics[clip,trim=1.8cm 1.1cm 1.7cm 1.07cm,width=0.48\textwidth]{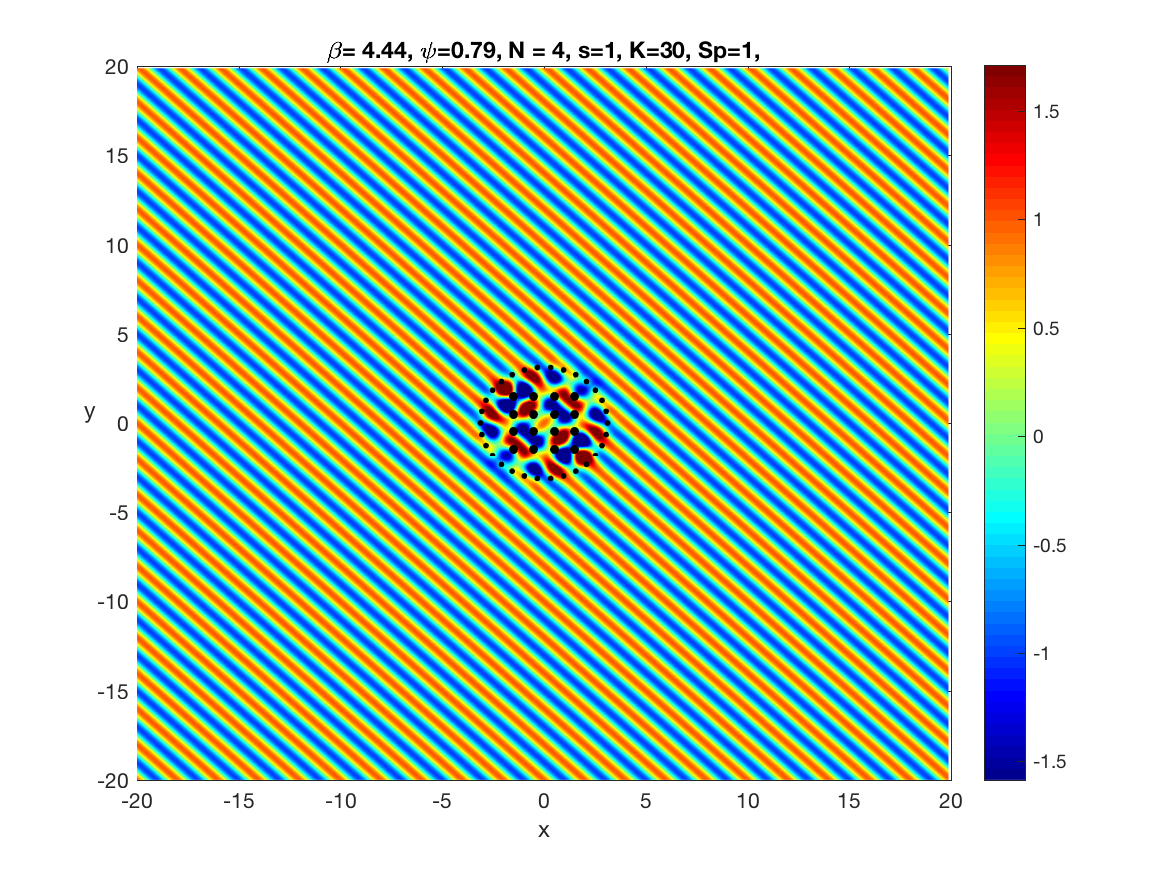}} \\
\hspace{-0.8cm}
\subfigure[]{
\includegraphics[clip,trim=1.0cm 0cm 1.7cm 1.1cm,width=0.445\textwidth]{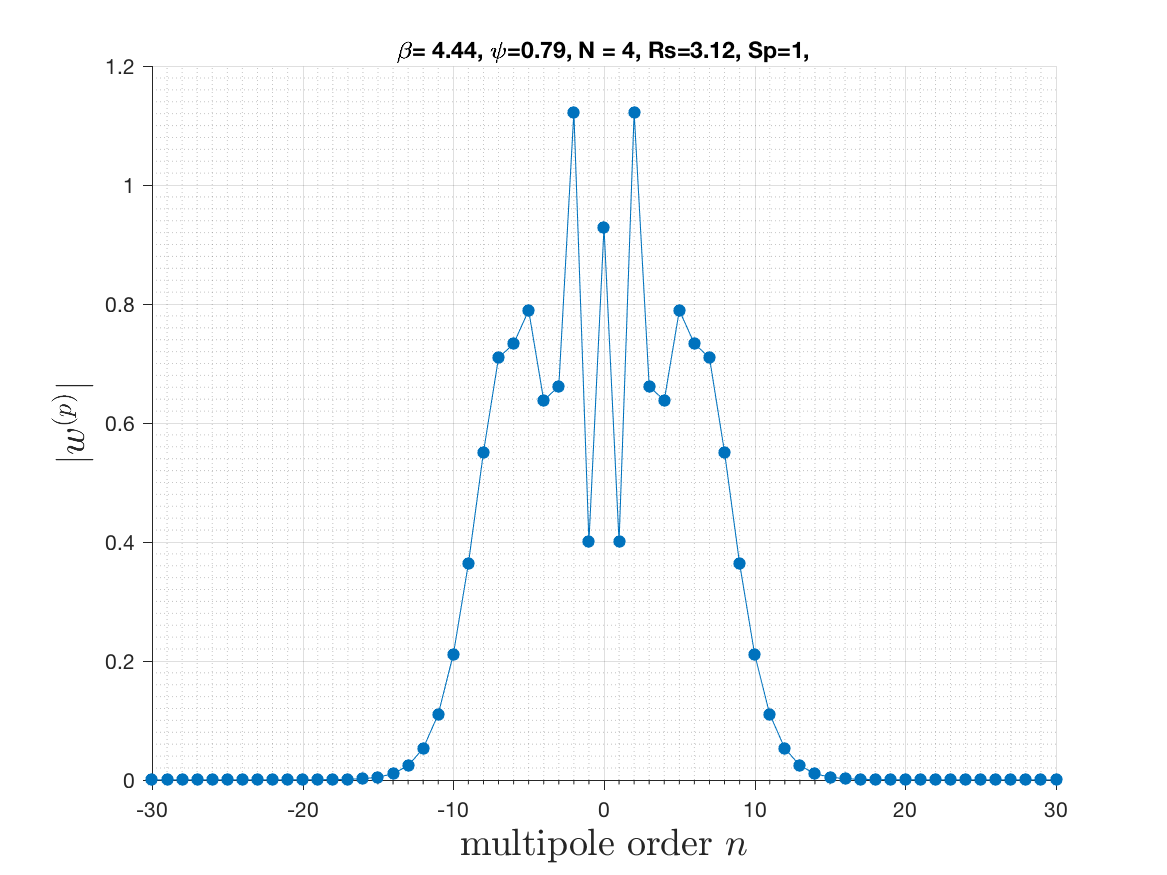}} ~~~~~~~ 
\subfigure[]{
\includegraphics[clip,trim=0.8cm 0cm 1.5cm 0.6cm,width=0.48\textwidth]{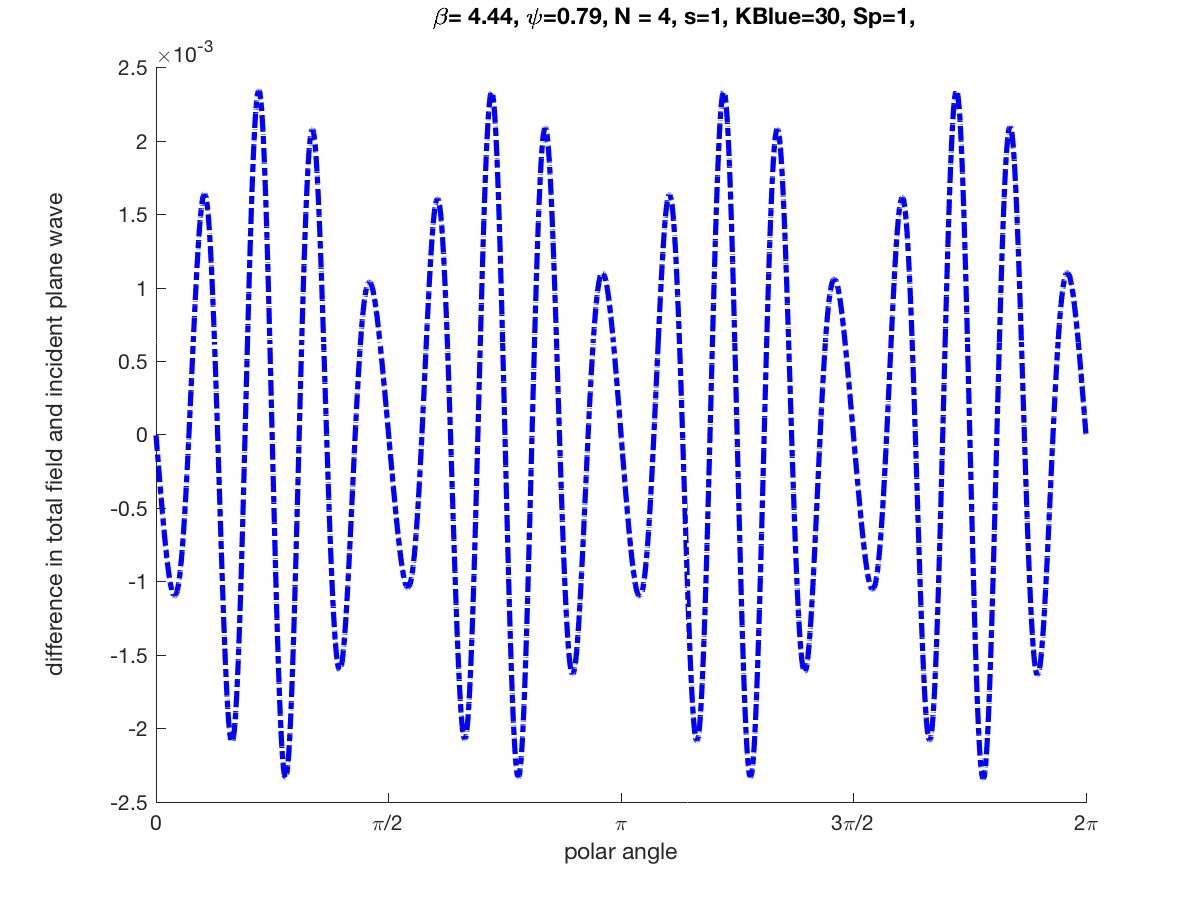}}
\caption{$4\times 4$ ($N=4$) cluster of pins with a plane wave incident at $\theta_i=\pi/4$ for $\beta=4.44$. Flexural wave amplitudes for (a) an uncloaked, (b) a cloaked cluster with $K=30$, (c)  $|w^p|$ versus multipole order $n$ for a circle of radius $a_s=(3\sqrt{2}+2)/2$, (d) absolute error evaluated on a circle of radius $15$ versus the polar angle.}
\label{beta_4p44_all}
\end{figure}

We finally consider $\beta=4.44$ for an angle of incidence $\theta_i=\pi/4$. In Figs.~\ref{bd_5bands} and \ref{3_disp_surf}(a) this value of the spectral parameter is in the vicinity of a Dirac-like point, where we expect to see some effects of dynamic neutrality within the finite cluster that shares the same periodicity as the infinite array in Figs.~\ref{bd_5bands}, \ref{3_disp_surf}. This is apparent in Fig.~\ref{beta_4p44_all}(a) where we see a similar pattern to that of Fig.~\ref{neut_M}(b). It is clear from Fig.~\ref{beta_4p44_all}(b) that we can once again achieve effective cloaking, and absolute error of the desired order is obtained for $30$ sources using Fig.~\ref{beta_4p44_all}(c).

\section{Concluding remarks}
\label{conc_rem}

We have presented a powerful mathematical method for cloaking finite clusters of pins from flexural waves in a Kirchhoff plate. The approach uses multipole methods to cancel wave scattering up to a specific order, through the use of active sources located exterior to the cluster. A system of algebraic equations is derived to determine the active source amplitudes required for effective cloaking. Here, we studied only the square array but the method is equally applicable to other Bravais lattices, such as hexagonal and rectangular arrays. 

We have demonstrated, through several illustrative examples, the capability of the cloaking algorithm for key values of the spectral parameter $\beta$, for which interesting features arise in the corresponding infinite array.  Two types of configuration of active sources were considered: positioning the active control sources on the grid's rows and columns adjacent to the cluster of pins, and sources located on a circle surrounding the cluster. The latter approach was adopted for higher frequencies at which the former approach became insufficient.

Locating the active sources on the circle provides greater flexibility in terms of their quantity and also enables us to identify precisely the order of multipoles to be annulled. This is accomplished by plotting the absolute value of the propagating part of the total field on this circle, at which the sources are located, versus the multipole order.  Thus, given a frequency and a finite array, we can predict how many sources are required to achieve a specified level of efficient cloaking. The results obtained for point scatterers, such as the rigid pins investigated here, may be readily extended to the case of finite-sized inclusions for which they act as a starting point.



\section*{Acknowledgements} 
We gratefully acknowledge the financial support from the EPSRC Programme Grant EP/L024926/1. J. O'Neill would also like to acknowledge the financial support from the EPSRC through the grant EP/L50518/1.


\begin{thebibliography}{9}



\bibitem{UL}
 U. Leonhardt,
Optical conformal mapping,
{\it Science}, {\bf 312}, (2006), 1777-1780.

\bibitem{JBP_DS_DRS}
J.B. Pendry, D. Schurig  and D.R. Smith,  
Controlling electromagnetic fields,
{\it Science}, {\bf 312}, (2006), 1780-1782.

 
\bibitem{GWM_MB_JRW}
G.W. Milton, M. Briane and J.R. Willis,
On cloaking for elasticity and physical equations with a transformation invariant form,
{\it New J. Phys.}, {\bf 8}, (2006), 248.

\bibitem{SG_RCMP_SE_ABM_MF_NAPN}
S. Guenneau, R.C. McPhedran, S. Enoch, A.B. Movchan, M. Farhat and N-A.P. Nicorovici,
The colours of cloaks, 
{\it J. Opt.}, {\bf 13}, (2011), 024014.


 \bibitem{NS_MW_MW}
N. Stenger,  M. Wilhelm and M. Wegener,
Experiments on elastic cloaking in thin plates,
{\it Phys. Rev. Lett.}, {\bf 108}, (2012), 014301.

\bibitem{NAPN_GWM_RCM_LCB}
N-A.P. Nicorovici, G.W. Milton, R.C. McPhedran and L.C. Botten,
Quasistatic cloaking of two-dimensional polarizable discrete systems by anomalous resonance,
{\it Opt. Express}, {\bf 15}, (2007), 6314-6323.

\bibitem{MF_SG_ABM_SE}
M. Farhat, S. Guenneau, A.B. Movchan and S. Enoch,
Achieving invisibility over a finite range of frequencies,
{\it Opt. Express}, {\bf 16}, (2008), 5656-61.

\bibitem{MF_SE_SG_ABM}
M. Farhat, S. Enoch, S. Guenneau and A.B. Movchan,
Broadband cylindrical acoustic cloak for linear surface waves in a fluid,
{\it Phys. Rev. Lett.}, {\bf 101}, (2008), 134501.

\bibitem{ISJ_MB_NVM_ABM}
I.S. Jones, M. Brun, N.V. Movchan  and A.B. Movchan,
Singular perturbations and cloaking illusions for elastic waves in membranes and Kirchhoff plates, 
{\it International Journal of Solids and Structures}, {\bf  69-70}, (2015),  498.

\bibitem{DJC_ISJ_NVM_ABM_MB_RCM}
D.J. Colquitt, I.S. Jones, N.V. Movchan, A.B. Movchan, M. Brun and R.C. McPhedran,
Making waves round a structured cloak: lattices, negative refraction and fringes,
{\it Proc. R. Soc. A}, {\bf 469}, (2013), 20130218. 

\bibitem{Alu_DoCloaks} F. Monticone and A. Al\`u, Do Cloaked Objects Really Scatter Less?
{\it Phys. Rev. X}, {\bf 3}, (2013), 041005. 

\bibitem{P-YC_CA_AA}
P-Y. Chen, C. Argyropoulos and A. Al\`u,
Broadening the cloaking bandwidth with non-Foster metasurfaces,
{\it Phys. Rev. Lett.}, {\bf 111}, (2013), 233001.

\bibitem{GVF_GWM_DO_2009a}
F. Guevara Vasquez, G.W. Milton and D. Onofrei,
Active exterior cloaking for the 2D Laplace and Helmholtz equations,
{\it Phys. Rev. Lett.}, {\bf 103}, (2009), 073901.

\bibitem{GVF_GWM_DO_2009b}
F. Guevara Vasquez, G.W. Milton  and D. Onofrei,
Broadband exterior cloaking,
{\it Optics Express}, {\bf 17}, (2009), 14800.

\bibitem{GVF_GWM_DO_2011}
F. Guevara Vasquez, G.W. Milton and  D. Onofrei,
Exterior cloaking with active sources in two dimensional acoustics,
{\it Wave Motion}, {\bf 48}, (2011), 515.

\bibitem{GVF_GWM_DO_PS}
F. Guevara Vasquez, G.W. Milton,  D. Onofrei and P. Seppecher, 
Transformation elastodynamics and active exterior acoustic cloaking,
{\it Acoustic Metamaterials, Springer Series in Materials Science}, {\bf 166}, (2013), 289-318.

\bibitem{JO_OS_RCM_ABM_NVM}
J. O'Neill, \"O. Selsil, R.C. McPhedran, A.B. Movchan and N.V. Movchan,
Active cloaking of inclusions for flexural waves in thin elastic plates, 
{\it Quart. J. Mech. Appl. Math.}, {\bf 68} (3), (2015), 263. 

\bibitem{JO_OS_RCM_ABM_NVMCHM}
J. O'Neill, \"O. Selsil, R.C. McPhedran, A.B. Movchan, N.V. Movchan and C. Henderson Moggach,
Active cloaking of resonant coated inclusions for waves in membranes and Kirchhoff plates, 
{\it Quart. J. Mech. Appl. Math.}, {\bf 69} (2), (2016), 115. 

\bibitem{ANN_FAA_WJP1}
A.N. Norris, F.A. Amirkulova and W.J. Parnell,  
Source amplitudes for active exterior cloaking,
{\it Inverse Problems}, {\bf 28}, (2012), 105002.

 \bibitem{ANN_FAA_WJP2}
A.N. Norris, F.A. Amirkulova and W.J. Parnell,
Active elastodynamic cloaking, 
{\it Mathematics and Mechanics of Solids}, {\bf 19},  (2013), 603.

\bibitem{GF_WJP_ANN}
G. Futhazar, W.J. Parnell and A.N. Norris,
Active cloaking of flexural waves in thin plates,
{\it Journal of Sound and Vibration}, {\bf 356}, (2015), 1.

\bibitem{NAPN_RCM_LCB}
N.-A.P. Nicorovici, R.C. McPhedran and L.C. Botten,
Relative local density of states and cloaking of finite clusters of coated cylinders,
{\it Wave Random Complex}, {\bf 21}(2), (2011), 248-277. 

\bibitem{Mace2}
B.R. Mace,
The vibration of plates on two-dimensionally periodic point supports, 
{\it J. Sound Vibration}, {\bf 192} (3), (1996), 629.

\bibitem{Mead}
D.J. Mead, 
Wave propagation in continuous periodic structures: Research contributions from Southampton, 1964-1995,
{\it J. Sound Vibration}, {\bf 190} (3), (1996), 495.

\bibitem{TARVC1}
T. Antonakakis and R.V. Craster,
High-frequency asymptotics for microstructured thin elastic plates and platonics,
{\it Proc. R. Soc. A}, {\bf 468} (2141), (2012), 1408. 

\bibitem{TARVCSG1}
T. Antonakakis, R.V. Craster and S. Guenneau, 
High-frequency homogenization of zero-frequency stop band photonic and phononic crystals,
{\it New J. Phys.}, {\bf 15}, (2013), 103014. 

\bibitem{TARVCSG2}
T. Antonakakis, R.V. Craster and S. Guenneau, 
Asymptotics for metamaterials and photonic crystals,
{\it Proc. R. Soc. A}, {\bf 469} (2152), (2013), 20120533.

\bibitem{TARVCSG3}
T. Antonakakis, R.V. Craster and S. Guenneau, 
Moulding and shielding flexural waves in elastic plates,
{\it EPL}, {\bf 105}, (2014), 54004.

\bibitem{ABM_NVM_RCM}
A.B. Movchan  N.V. Movchan and R.C. McPhedran, 
Bloch-Floquet bending waves in perforated thin plates, 
{\it Proc. R. Soc. A}, {\bf{463}}, (2007), 2505-2518.

\bibitem{RCM_ABM_NVM}
R.C. McPhedran, A.B. Movchan and N.V. Movchan,
Platonic crystals: Bloch bands, neutrality and defects,
{\it Mech. Mater.}, {\bf 41}, (2009), 356. 

\bibitem{RCM_ABM_NVM_MB_MJAS}
R.C. McPhedran, A.B. Movchan, N.V.  Movchan, M. Brun and M.J.A. Smith,  
`Parabolic' trapped modes and steered Dirac cones in platonic crystals,
{\it Proc. R. Soc. A}, {\bf 471}, (2015), 20140746.

\bibitem{NVM_RCM_ABM_CGP}
N.V. Movchan, R.C. McPhedran, A.B. Movchan and C.G. Poulton,
Wave scattering by platonic grating stacks,
{\it Proc. R. Soc. A}, {\bf 465}, (2009), 3383. 


\bibitem{CGP_RCM_NVM_ABM}
C.G. Poulton, R.C. McPhedran, N.V.  Movchan and A.B. Movchan,
Convergence properties and flat bands in platonic crystal band structures using the multipole formulation, 
{\it Wave Random Complex}, {\bf 20} (4),  (2010), 702. 

\bibitem{SGH_NVM_ABM_RCM1}
S.G. Haslinger, N.V. Movchan, A.B. Movchan and R.C. McPhedran,
Transmission, trapping and filtering of waves in periodically constrained elastic plates,
{\it Proc. R. Soc. A}, {\bf 468}, (2012), 76. 

\bibitem{SGH_NVM_ABM_RCM2}
S.G. Haslinger, N.V. Movchan, A.B. Movchan and R.C. McPhedran,
Symmetry and resonant modes in platonic grating stacks, 
{\it Wave Random Complex}, {\bf 24} (2), (2014)126. 

\bibitem{DVE_RP}
D.V. Porter and R. Porter, 
Penetration of flexural waves through a periodically constrained thin elastic plate in {\it vacuo} and floating on water,
{\it J. Eng. Math.}, {\bf 58}, (2007), 317.

\bibitem{CML_PAM}
C.M. Linton and P.A. Martin,
Semi-infinite arrays of isotropic point scatterers. A unified approach,
{\it SIAM J. Appl. Math.}, {\bf 64}(3), (2004), 1035-1056.

\bibitem{LLF}
L.L. Foldy,
The multiple scattering of waves,
{\it{Phys. Rev}}, {\bf{67}}, (1945), 107-119. 

\bibitem{LL_JDJ_MS}
L. Lu, J.D. Joannopoulos and M. Solja\^{c}i\'{c},
Topological photonics, {\it Nat. Photonics}, {\bf 8}, (2014), 821. 


\bibitem{AHCN_FG_NMRP_KSN_AKG}
A.H. Castro Neto, F. Guinea, N.M.R. Peres, K.S. Novoselov, and A.K. Geim,
The electronic properties of graphene, 
{\it Rev. Mod. Phys.}, {\bf  81}, (2009), 109.

\bibitem{MJAS_RCM_MHM}
M.J.A. Smith, R.C. McPhedran and M.H. Meylan, 
Double Dirac cones at $\kappa$ = 0 in pinned platonic crystals,
{\it Waves in Random and Complex Media}, {\bf 24}(1), (2014), 35-54.

\bibitem{JM_YW_CTC_ZQZ}
J. Mei, Y. Wu, C.T. Chan and Z-Q. Zhang,
First-principles study of Dirac and Dirac-like cones in phononic and photonic crystals,
{\it Phys. Rev. B}, {\bf 86}, (2012), 035141.


\bibitem{RVC_JK_AVP}
R.V. Craster, J. Kaplunov and A.V. Pichugin,
High-frequency homogenization for periodic media, {\it Proc. Roy. Soc. A} {\bf 466}, (2010), 2341-2362.

\bibitem{RVC_JK_JP}
R.V. Craster, J. Kaplunov and J. Postnova,
High-frequency asymptotics, homogenisation and localisation for lattices, {\it Q. J. Mech. Appl. Math.} {\bf 63}, (4), (2010), 497-519.

\bibitem{SGH_RVC_ABM_NVM_ISJ}
S.G. Haslinger, R.V. Craster, A.B. Movchan, N.V. Movchan and I.S. Jones,  Dynamic interfacial trapping of flexural waves in structured plates. {\it. Proc. Roy. Soc. A} {\bf 472}, (2015), 20150658. 

\end{thebibliography}
\end{document}